\documentclass[fleqn,usenatbib]{mnras}

\usepackage[T1]{fontenc}

\DeclareRobustCommand{\VAN}[3]{#2}
\let\VANthebibliography\thebibliography
\def\thebibliography{\DeclareRobustCommand{\VAN}[3]{##3}\VANthebibliography}

\usepackage{comment}
\usepackage{caption}

\usepackage{graphicx}	
\usepackage{amsmath}	
\usepackage{amssymb}	
\usepackage{xcolor}
\usepackage{multirow}

\usepackage{soul}

\usepackage[flushleft]{threeparttable}

\usepackage{newtxtext,newtxmath}

\title[Cosmology from Planck and ACT SZ clusters]{Cosmological inference from combining Planck and ACT cluster counts}

\author[E. Lee et al.]{
Eunseong Lee,$^{1,2,3}$\thanks{E-mail: eunseong@sas.upenn.edu}
Richard Battye,$^{1}$\thanks{E-mail: richard.battye@manchester.ac.uk}
Boris Bolliet $^{1,4,5}$
\\
$^{1}$Jodrell Bank Centre for Astrophysics, School of Physics and Astronomy, University of Manchester, Oxford Road, Manchester M13 9PL, U.K.\\
$^{2}$Department of Physics, Cornell University, Ithaca, NY 14853, U.S.A. \\
$^{3}$Department of Physics and Astronomy, University of Pennsylvania, Philadelphia, PA 19104, U.S.A. \\
$^{4}$Kavli Institute for Cosmology, University of Cambridge, Madingley Road, Cambridge CB3 0HA\\
$^{5}$DAMTP, Centre for Mathematical Sciences, Wilberforce Road, Cambridge CB3 0WA, U.K.
}
\date{Accepted XXX. Received YYY; in original form ZZZ}

\pubyear{2024}

\begin{document}
\label{firstpage}
\pagerange{\pageref{firstpage}--\pageref{lastpage}}
\maketitle

\begin{abstract}
We have adapted the {\it Planck} cluster likelihood in such a way that it can be applied to the sample of clusters detected by the Atacama Cosmology Telescope (ACT). Applying it to the 2016 sample from {\it Planck} and the 2018 sample from ACT we find, by fixing the cosmology using CMB observations and the cluster model adopted by {\it Planck}, that the mass bias required by the two are $1-b_{\rm Planck}=0.61\pm 0.03$ and $1-b_{\rm ACT}=0.75\pm 0.06$. These are broadly in agreement but hint that the model could be adapted to reach a better agreement. By normalizing the cluster model using weak lensing observations, we find evidence for either evolution in the cluster model, quantified by the cluster modeling parameter describing redshift dependence $\beta=0.86 \pm 0.07$ using an updated CCCP-based normalization, or evolution in the cosmological model quantified by the dark energy equation of state parameter $w=-0.82 \pm 0.07$. 
\end{abstract}




\section{Introduction} 

Galaxy clusters form at the extreme peaks of the cosmological mass function. They have long been used to probe cosmological parameters~\citep{Allen2011} and it has been shown that their abundance can be predicted from quasi-linear theory using a mass function measured in numerical simulations~\citep{Jenkins2001, Tinker2008}. From this it is easy to see that number of clusters as a function of mass and redshift is sensitive to the present amplitude of density fluctuations, $\sigma_{8}$, and the cosmic matter density relative to critical, $\Omega_{\rm m}$, which are directly related to the growth of structures in the Universe, as well as a number of other cosmological parameters including the dark energy equation of state parameter, $w$, \citep[see, e.g.,][for a detailed description of the basic methodology]{Battye:2003bm} and the sum of neutrino masses, $\Sigma m_{\nu}$. Cosmological constraints from cluster counts are complementary to those which reflect different epochs in the Universe's history, such as Cosmic Microwave Background (CMB) anisotropies~\citep{Planck2018vi, act2020aiola, sptCMB2022} and Baryon Acoustic Oscillations (BAO)~\citep{de_Mattia_2020, Neveux_2020, Bautista_2020}.

Clusters were first identified using optical observations~\citep{Abell1958}, but most cosmological studies have typically focused on their detection of the hot intracluster medium (ICM) firstly by bremsstrahlung radiation into X-rays~\citep[see, e.g.,][]{Borgani2001,Reiprich_and_Bohringer2002,Vikhlinin:2008ym,Mantz2010,2012arXiv1209.3114M}, and more recently using the Sunyaev-Zeldovich (SZ) effect~\citep{Sunyaev:1970eu,Sunyaev:1972eq} which is due to the inverse Compton scattering of CMB photons. In addition, large-scale photometric surveys of the Universe in the optical-near infrared wavebands have prompted attempts to do this using the clustering properties of galaxies~\citep{DES:2020cbm, DES2022}.

The focus of this paper is cosmological studies using cluster detection via the SZ effect which results in temperature decrements below the null frequency $\sim 217\,{\rm GHz}$ and increments above it. The surface brightness does not depend on the redshift of the cluster, but only on the integrated gas pressure within the cluster along the line of sight, allowing us to detect more cluster candidates at higher redshifts. Moreover, the amplitude of the SZ effect is shown to be a low scatter proxy for the mass of the cluster compared to the X-ray flux \citep[see, e.g.,][for a detailed review]{Birkinshaw:1998qp}. 

The present state-of-the-art galaxy cluster catalogues have come from large area surveys using the {\it Planck} space telescope~\citep{planck2014source,Planck2014xx,Planck2016source,Planck2016xxiv}, the Atacama Cosmology Telescope (ACT)~\citep{Hasself2013,Hilton2017,Hilton2021} and South Pole Telescope (SPT)~\citep{Vanderlinde2010,Bleem2015,deHaan2016,Bocquet2018,Huang2020,Salvati:2021gkt,bleem2023}. These have yielded substantial galaxy cluster samples over a thousand tSZ cluster candidates with redshift measurements and they have been used to constrain $\sigma_8$ and $\Omega_{\rm m}$. Of key importance is the normalization of the mass-observable relation which is often quantified in terms of a mass bias $1-b$ \citep[see, e.g.,][for a discussion of the rationale behind this]{Planck2014xx}. The fact that the mass of a cluster is not directly measurable makes the determination of underlying mass inferred from the observable crucial in the prediction of cluster abundance. On that account, the values of mass biases obtained from weak lensing mass measurements can be useful as gravitational lensing is a direct probe of the total mass of the cluster~\citep{Linden_2014,Hoekstra2015,Melin2015,Herbonnet2019cccp,Zubeldia2019}. 

The goal of this paper is to compare two recent SZ cluster surveys; {\it Planck} and ACT with the hope of combining them in order to improve the power of cosmological inference.  The paper is structured as follows.
In Section~\ref{sec:methodology}, we review the details of recent {\it Planck} and ACT analyses; how the observable-mass scaling relations are defined, how the survey completeness is determined, and how the cluster counts likelihood functions are built. In Section~\ref{sec:external}, we describe the external data we introduce for cosmological analyses. Noting that the two likelihood functions are slightly different, but fundamentally the same, in Section~\ref{ss:result_each}, we compare their conclusions using the cluster models adopted in the individual analyses before and in Section~\ref{ss:combined}, on the back of adequate agreement between the two, we adopt the cluster model used in the {\it Planck} analysis to describe both. We find that the wide redshift range probed by the combined sample allows a wider range of cluster models and cosmological parameters to be probed. 
We summarise and discuss the results in Section~\ref{ss:conc}.
We note that a similar comparison was attempted by \citealp{Salvati:2021gkt} for {\it Planck} and SPT. All uncertainties are given at the 68 percent confidence level.
Throughout the paper, we use $\ln$ for the natural log and $\mathrm{log}$ for the base 10 logarithm. We note that the total neutrino mass is fixed to $0.06 \mathrm{eV}$ and in our analyses, we treat massive neutrinos in the same way as {\it Planck}. For detailed neutrino treatment in cluster counts analyses we refer \citealp{Bolliet:2019zuz}.

\section{Methodology} \label{sec:methodology}
In order to constrain cosmological parameters we need to develop a likelihood to compare number count predictions against observed data. Here, we review the data and its modeling and advance the cluster counts likelihood analyses used in recent SZ cluster studies of {\it Planck} and ACT whose details are summarised in table~\ref{table:comp_plc_act}. Ultimately, all these works have dealt with the same issues; how to select the cluster sample, how to model the ICM, and how to connect observed flux to mass.
Assuming a unified framework for clusters, we show that the mass-observable scaling relations used in {\it Planck} and ACT can be connected and hence interchanged.

\begin{table*}
\centering
\makebox[\textwidth]{
\begin{tabular} {l c c}
    \hline
	Experiment & Planck & ACT \\ 
	\hline
	Number of clusters & 439 & 182 (59) \\
	SNR threshold & 6 & 4 (5.6) \\
	Sky coverage [deg$^{2}$] & 41253$^{\rm a}$ (entire sky) & 987.5 (E-D56) \\
	Frequency [GHz] & 100, 143, 217, 353, 545, 857 & 148 \\
	Resolution [arcmin] & 5 $\sim$ 10 & 1.4 \\
	Redshift range & [0, 1] & [0.1, 1.4] \\
	Mean SZ mass [$10^{14} M_{\odot}$] & 5.4 & 3.1 \\ 
	Detection method & Multi-freq matched filter (MMF3) & Single-freq matched filter \\
	Cluster physics & UPP & UPP (+ B12, adiabatic, nonthermal20) \\
	SZ observable & Y$_{500}$ (integrated) & y$_{0}$ (peak) \\
    Mass scaling & X-ray mass & X-ray, dynamical mass \\ 
    Likelihood & Binned in redshift and SNR & Unbinned (individual) \\
	Priors & WtG, CCCP, CMBlens & None \\ 
	Analysis reference & P16 & Ha13 \\
	Catalogue reference & P16 & Hi18 \\
	\hline
\end{tabular}}
\vspace{0.5em}
\caption{Observational data details. For the number of clusters and SNR threshold, the values corresponding to the cosmological sample are shown in the parenthesis for ACT. $^{\rm (a)}$After masking, the survey area is reduced to 65$\%$ of the sky (\protect\citealp{Planck2016source}).} 
\label{table:comp_plc_act}
\end{table*}

\subsection{Planck cluster likelihood}
\begin{figure}
\centering
\includegraphics[width=\columnwidth]{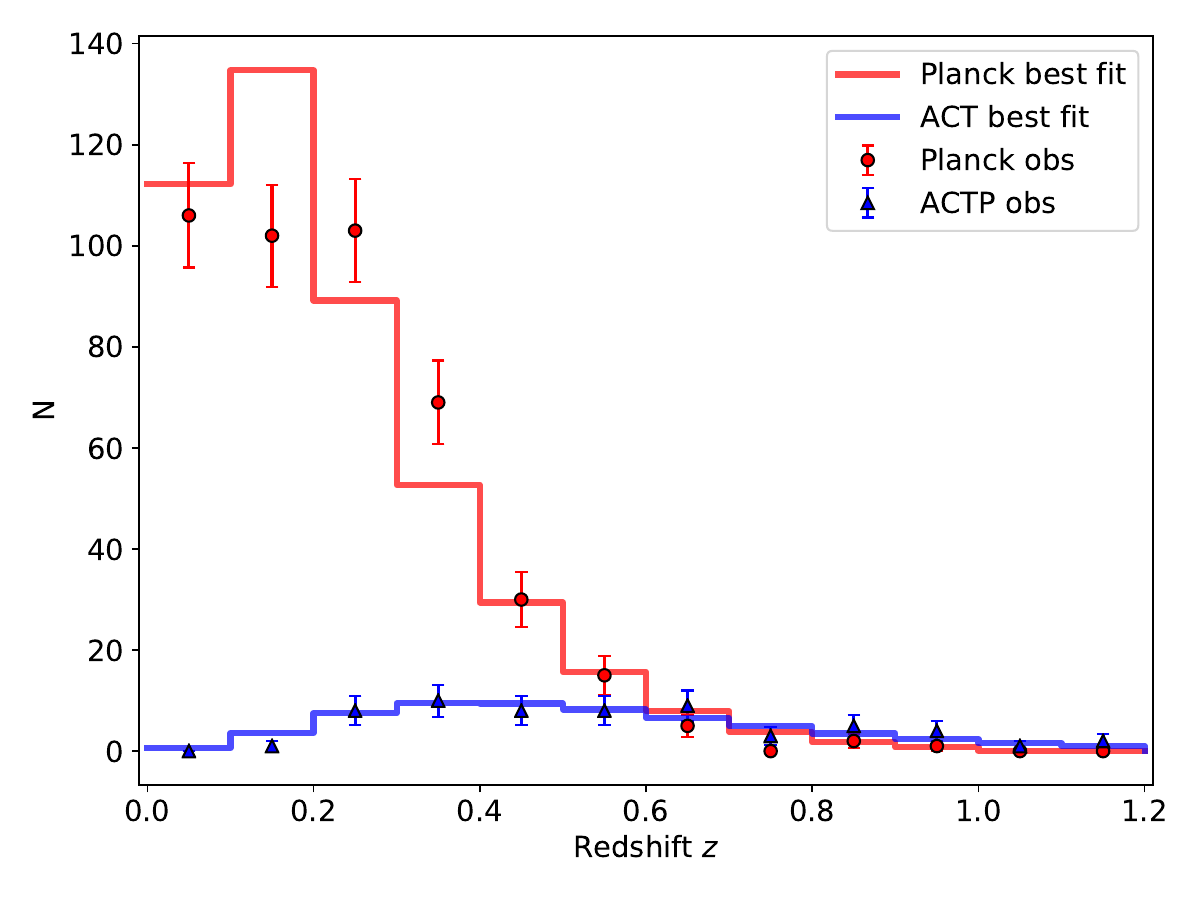}
\includegraphics[width=\columnwidth]{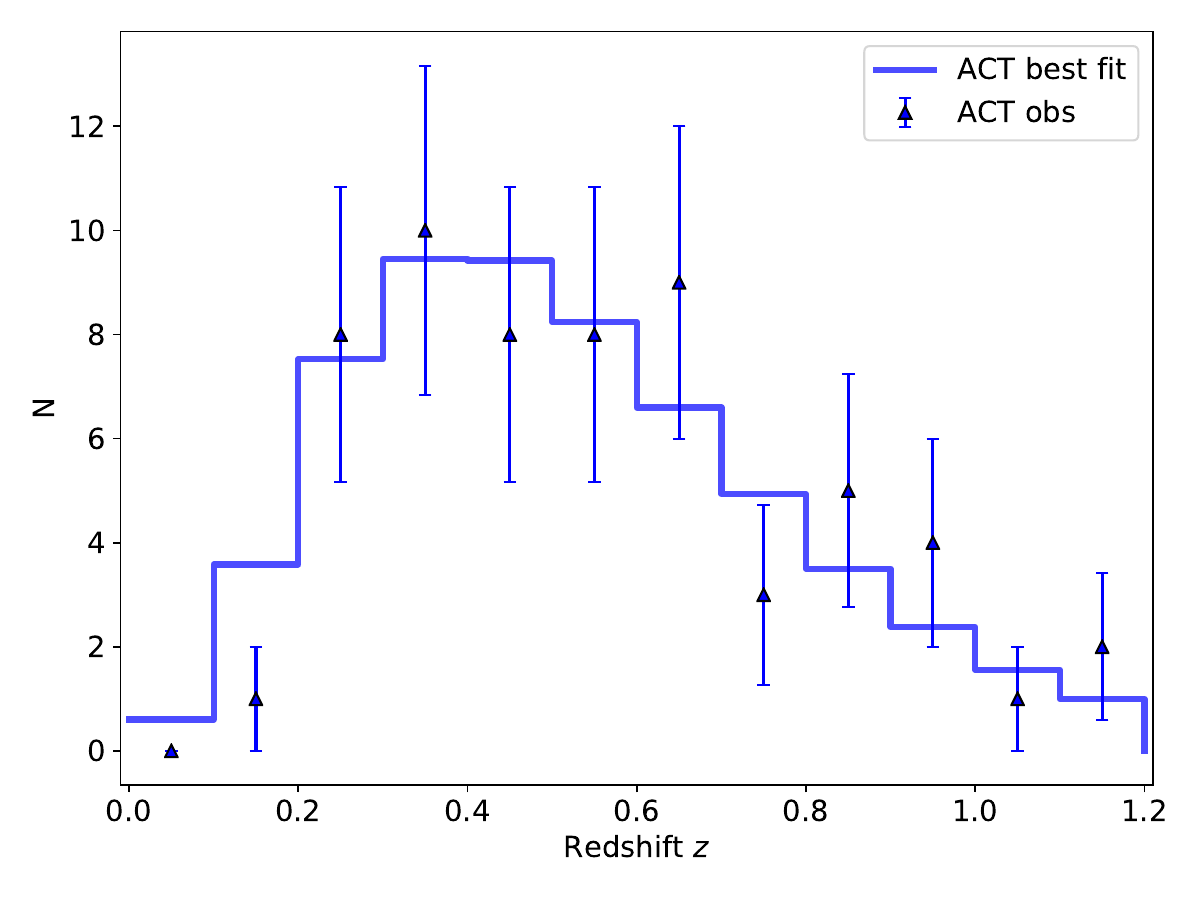}
\caption{Cluster distribution of {\it Planck} and ACT cosmological sample as a function of redshift. The size of the error bar is given by assuming a Poissonian distribution in each redshift bin. {\it Planck} MMF3 sample consists of 433 clusters with confirmed redshift above SNR cut of 6 \protect\citep{Planck2016source} and 59 ACT high confidence sample is given by SNR$_{2.4}$ cut of 5.6 \protect\citep{Hilton2017}. 6 additional clusters without redshift information are included by renormalizing the observed number counts in {\it Planck} analysis. We see that the {\it Planck} detects clusters at low redshifts and ACT, typically, detects the ones in a higher redshift range. The observed cluster counts are shown as points with error bars and are compared to predictions of the best-fit models which are shown as solid lines. {\it Planck} best-fit model (448) and ACT best-fit model (58) are obtained from the baseline likelihood using updated external datasets, which will be explained in detail in Section~\ref{ss:result_each}. Two-dimensional likelihood over redshift and signal-to-noise is used for both experiments.}
\label{fig:Nz_hist}
\end{figure}

The European Space Agency's {\it Planck} space telescope scanned the entire sky numerous times between 2009 to 2013 and clusters were detected through the SZ spectral signature across the six highest frequency bands\footnote{The High-Frequency Instrument (HFI) of {\it Planck} covers frequency bands centred at 100, 143, 217, 353, 545, and 857 GHz.}. Since the angular resolution of {\it Planck} (5-10 arcmin) at relevant frequencies is rather large compared to that of ACT ($\sim$ 1 arcmin), {\it Planck} is able to detect more massive, larger angular size clusters at lower redshift. In addition the {\it Planck} data benefits from the full sky coverage and the lack of atmosphere noise.  The {\it Planck} cosmological sample from the initial catalogue \citep[PSZ,][]{planck2014source} contains 189 candidates defined by a signal-to-noise ratio (hereafter, SNR or $q$) cut of 7 and the full-mission catalogue \citep[PSZ2,][]{Planck2016source} contains 439 candidates defined by $q>6$. 

Connecting what is observed to the mass of the cluster is a crucial part of this work. The observable specified in the {\it Planck} catalogue is the integrated SZ flux $Y_{500}$ within the angular size $\theta_{500}$. A corresponding mass can be defined as $M_{500}$, which is a total mass within a sphere of radius $R_{500}$ that encloses the over-density of 500 times the critical density of the Universe.  
 
The {\it Planck} mass-observable scaling relation is given as in \citealp{Planck2016xxiv} (hereafter, P16) by 
\begin{equation}\label{eq:plc_Y500}
	\overline{Y}_{500} = 
	Y_{*}\left[\frac{h}{0.7}\right]^{\alpha-2}
	\left[\frac{(1-b)M_{500}}{6\times 10^{14}M_{\odot}}\right]^{\alpha} 
	[E(z)]^{\beta} \left[\frac{D_{\!A}(z)}{\text{Mpc}}\right]^{-2} \!\! 10^{-4}\,,
\end{equation}
and the angular size of the cluster is 
\begin{equation}\label{eq:plc_theta500}
	\overline{\theta}_{500} = \theta_{*}\left[\frac{h}{0.7}\right]^{-\frac{2}{3}}  \left[\frac{(1-b)M_{500}}{3\times 10^{14}M_{\odot}}\right]^{\frac{1}{3}} [E(z)]^{-\frac{2}{3}}\left[\frac{D_{\!A}(z)}{500\,\text{Mpc}}\right]^{-1}\,,
\end{equation}
where the dimensionless Hubble parameter is defined by $E(z) = \sqrt{\Omega_{m}(1+z)^{3}+\Omega_{\Lambda}}$ with the total matter density, $\Omega_{\mathrm{m}}$, and a cosmological constant, $\Omega_{\Lambda}$, and $D_{\!A}(z)$ is the angular diameter distance to redshift $z$ which relates an observed angular size to a physical radius of the cluster through $\theta_{500} = R_{500} / D_{\!A}(z)$. The normalisation factor for $\theta_{500}$ is given by $\theta_{*} = 6.997$ arcmin and the scaling parameters of $Y_{500}$ are $Y_{*}$, $\alpha$, and $\beta$, which describe the normalization, mass dependence, and redshift evolution of the relation, respectively. The {\it Planck} SZ mass is defined as $M_{\text{plc}} = (1-b)M_{500}$ with the mass bias parameter $1-b$ which is presumed to account for all the possible uncertainties regarding the true cluster mass $M_{500}$. It is possible that the mass bias could be a function of mass and redshift, however, for the moment it will be assumed to be a constant. The values of scaling parameters are fitted from the X-ray observation of the {\it Planck} clusters and they are summarised in Table~\ref{table:scaling_params}. {\it Planck} observable-mass scaling relation is derived in two steps. Firstly, they calibrated the hydrostatic mass $(1-b)M_{500}$ against the X-ray mass observable $Y_{\mathrm{X}}$ using 20 low redshift clusters, which were provided by \citealp{Arnaud2010} (hereafter, A10).
The X-ray mass proxy is computed iteratively from the observed temperature and gas mass profile with respect to the radius following the method in \citealp{kravtsov2006}. As a second step, the observed SZ signal within the radius of $R_{500}$ obtained from the position of the X-ray peak is calibrated with the X-ray mass measurements using 71 clusters in {\it Planck} cosmological sample. Scaling parameters for slope and normalisation were fitted to $Y_{500}$ and $M_{500}^{Y_{\mathrm{X}}}$ relation from A10 and corrected for Malmquist bias. Uncertainty about combining two scaling relations was added assuming that the scatter between two relations is uncorrelated.
Determination and assessment of {\it Planck} scaling parameter values are described in detail in Appendix A in \citealp{Planck2014xx} (hereafter, P14). 

The number of clusters as a function of redshift $z$ can be predicted as
\begin{equation}\label{eq:plc_1d_dNdz}
	\frac{{\rm d}N}{{\rm d}z} = \Delta\Omega \, \frac{{\rm d}V}{{\rm d}z {\rm d}\Omega}(z) \int \!\! {\rm d}M_{500} \frac{{\rm d}n(z,M_{500})}{{\rm d}M_{500}} \, \chi(z,M_{500},\underline{x})\,,
\end{equation}
where $\Delta\Omega$ is the angular sky coverage of the survey, which is approximately 65\% of the sky for {\it Planck}, ${\rm d}V/({\rm d}z {\rm d}\Omega)$ is the comoving volume element and ${\rm d}n/{\rm d}M_{500}$ is the number of objects with mass $M_{500}$ and redshift $z$ per comoving volume, which is often called the halo mass function. Typically this is computed from N-body simulations and we use the mass function from \citealp{Tinker2008} with the overdensity of 500 times of the critical density.

It is necessary to take into account the uncertainty in the observed cluster quantities and the survey selection function. 
The survey completeness is defined as a probability distribution
\begin{equation}\label{eq:plc_1d_complt}
    \chi = \int \!\! {\rm d}\!\ln Y_{500} \int \!\!  {\rm d}\theta_{500} \, \text{P}(\ln Y_{500},\theta_{500}\,|\,z, M_{500}) \, \chi_{\textrm{erf}}(Y_{500},\theta_{500}, \underline{x})\,,
\end{equation}
where $\text{P}(\ln Y_{500},\theta_{500}\,|\,z, M_{500})$ is the product of a Gaussian distribution for the SZ flux in natural logarithm, $\ln Y_{500}$, centred on its predicted value, $\ln \overline{Y}_{500}$, of a cluster with mass $M_{500}$ and redshift $z$ and a delta function for the cluster of the angular size of $\theta_{500}$ around its predicted value $\overline{\theta}_{500}(z, M_{500})$, which can be written as
\begin{equation}\label{eq:plc_1d_scatter}
\begin{split}
    \text{P} &= \frac{1}{\sqrt{2\pi}\sigma_{\ln Y_{500}}} \exp \left[-\left( \frac{\ln Y_{500} - \ln \overline{Y}_{500} (z, M_{500})}{\sqrt{2}\sigma_{\ln Y_{500}}} \right)^{2} \right] \\
    & \quad \times \delta[\theta_{500} - \overline{\theta}_{500}(z, M_{500})]\,.
\end{split}
\end{equation}
Here, $\sigma_{\ln Y_{500}}$ is the estimated intrinsic scatter derived from the calibration of the $Y_{500}$-$M_{500}$ relation (see Appendix A in P14).  In principle, one could detect all clusters above a given mass limit depending on the survey. Since the number of clusters is a very steep function of mass, with a scatter around the mass limit, there are more clusters scattered up randomly from a lower mass than scattered down from above the mass limit. For a given set of cosmological parameters, the inclusion of intrinsic scatter will increase the total number of predicted clusters. The predicted values $\overline{Y}_{500}$ and $\overline{\theta}_{500}$ are given by the scaling relations in Eq.~\eqref{eq:plc_Y500} and Eq.~\eqref{eq:plc_theta500}.

Assuming a Gaussian probability distribution for the observed $Y_{500}$, 
\begin{equation}\label{eq:plc_1d_erf}
    \chi_{\textrm{erf}} (Y_{500},\theta_{500}, \underline{x}) = \frac{1}{2} \left[1 + \text{erf} \left( \frac{Y_{500} - q_{\textrm{cut}}\,\sigma_{Y_{500}}(\theta_{500}, \underline{x})}{\sqrt{2}\sigma_{Y_{500}}(\theta_{500}, \underline{x} )}\right) \right]\,,
\end{equation}
where $q_{\textrm{cut}}$ is an SNR threshold of the sample and a Gaussian uncertainty $\sigma_{Y_{500}}$ is the noise measurement for a filter size $\theta_{500}$ at a given location $\underline{x}$ in the map, which is derived as part of the cluster detection process. $\chi_{\textrm{erf}}$ describes how likely it is to detect a cluster of the SZ flux $Y_{500}$ and the size $\theta_{500}$ at a location $\underline{x}$ in the sky when the threshold of SNR is $q_{\textrm{cut}}$. This quantity does not depend on the cosmology as it is only related to the observed properties. This semi-analytic approach based on the error function is cross-validated with the Monte-Carlo method using simulated cluster injection into real sky maps (see Section 3.2 in P14 for a detailed description).

With a larger sample size from PSZ2, it was possible to use the cluster counts in two dimensions with respect to redshift $z$ and signal-to-noise $q$ following P16.  The number of clusters in each redshift bin in Eq~\eqref{eq:plc_1d_dNdz} is computed by integrating over signal-to-noise as follows 
\begin{equation}\label{eq:plc_2d_dNdzdq}
	{\frac{{\rm d}N}{{\rm d}z}}(q_{\rm min} < q < q_{\rm max}) = \int_{q_{{\rm min}}}^{q_{{\rm max}}} {\rm d}q \frac{{\rm d}N}{{\rm d}z {\rm d}q}\,,
\end{equation}
where ${\rm d}N\!/({\rm d}z {\rm d}q)$ is the distribution of clusters in redshift and signal-to-noise.
Then the survey completeness in Eq~\eqref{eq:plc_1d_complt} for ${\rm d}N\!/({\rm d}z {\rm d}q)$ becomes 
\begin{equation}\label{eq:plc_2d_complt}
  \chi_{_\mathrm{2D}} (z,M_{500},\underline{x}) = \int {\rm d}q \, \text{P}[q|\,\overline{q}_{\text{m}}(z,M_{500},\underline{x})]\,,
\end{equation}
where the predicted $\overline{q}_{\text{m}}$ is defined by the predicted SZ signal $\overline{Y}_{500}$ in Eq~\eqref{eq:plc_Y500} and the detection noise $\sigma_{\textrm{f}}$ at an expected filter scale of $\overline{\theta}_{500}$ in Eq.~\eqref{eq:plc_theta500},
\begin{equation}\label{eq:plc_2d_qmean}
    \overline{q}_{\text{m}} (z,M_{500},\underline{x}) \equiv \frac{\overline{Y}_{500}(z, M_{500})}{\sigma_{\textrm{f}}\big[\overline{\theta}_{500}(z,M_{500}), \underline{x}\big]}\,,
\end{equation}
and the completeness $\chi_{_\mathrm{2D}}$ describes the probability distribution of the observed SNR $q$ given the predicted $\overline{q}_{\text{m}}$ for a cluster at a sky location $\underline{x}$. As in the one-dimensional case of Eq.~\eqref{eq:plc_1d_complt}, this quantity consists of a term responsible for the effect of intrinsic scatter and another accounting for the survey selection function
\begin{equation}\label{eq:plc_2d_P}
    \text{P} (q|\overline{q}_{\text{m}}) = \!\! \int \!\! {\rm d} \!\ln q_{\text{m}} \, \text{P}[\ln q_{\text{m}}|\ln \overline{q}_{\text{m}}(z,M_{500},\underline{x})] \, \chi_{_\textrm{erf,2D}} (Y_{500},\theta_{500},\underline{x})\,,
\end{equation}
where the first term represents the intrinsic scatter in scaling relation in terms of the predicted $\overline{q}_{\text{m}}$ given by the predicted signal $\overline{Y}_{500} (z,M_{500})$ and the filter noise $\sigma_{\textrm{f}}(\overline{\theta}_{500}, \underline{x})$
\begin{equation}\label{eq:plc_2d_scatter}
	\text{P} (\ln q_{\text{m}}|\ln \overline{q}_{\text{m}})
	= \frac{1}{\sqrt{2\pi}\sigma_{\ln Y_{500}}}\exp \left[-\left( \frac{\ln q_{\text{m}} - \ln \overline{q}_{\text{m}} (z,M_{500},\underline{x})}{\sqrt{2}\sigma_{\ln Y_{500}}} \right)^{2}\right]\,,
\end{equation}
and the second term can be expressed as
\begin{equation}\label{eq:plc_2d_erf}
    \chi_{_\textrm{erf,2D}} (Y_{500},\theta_{500},\underline{x}) = \frac{1}{2} \left[1 - \text{erf} \left( \frac{q_{\textrm{cut}} - q_{\textrm{m}}(Y_{500},\theta_{500},\underline{x})}{\sqrt{2}}\right) \right]\,,
\end{equation}
which effectively describes a Gaussian probability distribution of the observed $q$ given the model $q_{\textrm{m}}$ in terms of the observed quantities with pure Gaussian noise.

The {\it Planck} likelihood function is constructed based on the observed number counts and the predicted number counts in bins for a given theoretical model. Assuming Poisson statistics \citep{Cash1979}, the probability of observing $N_{ij}$ clusters in $i$-th redshift and $j$-th SNR bin, given a theoretically expected number of $\overline{N}_{ij}$ in the same bin, is 
\begin{equation}\label{eq:plc_2d_lkl}
	\ln \mathcal{L} = \ln \mathcal{P} (N_{ij} | \overline{N}_{ij}) = \sum \limits_{i}^{N_{z}} \sum \limits_{j}^{N_{q}} \left[ N_{ij} \ln \overline{N}_{ij} - \overline{N}_{ij} - \ln(N_{ij}!) \right]\,,
\end{equation}
where $N_{z}$ and $N_{q}$ are the total number of redshift and signal-to-noise bins, respectively.
The PSZ2 sample is split into 10 redshift bins with $\Delta z=0.1$ and 5 SNR bins with $\Delta\log q = 0.25$ so that
\begin{equation}
    \overline{N}_{ij} = \frac{{\rm d}N}{{\rm d}z {\rm d}q}(z_i, q_j)\, \Delta z \Delta q\,.
\end{equation}
Note that the bins without observed clusters are included and the bins are assumed to be uncorrelated for simplicity. The posterior likelihood of cosmological parameters is computed based on the Bayes theorem.\footnote{We do not take into account the super sample covariance (\citealp{Payerne:2024lrv}) in the likelihood. We only consider the binned likelihoods here. \citep[see, e.g.,][for further details on other likelihood formalisms for cluster count cosmology.]{Payerne:2022alz, Zubeldia:2024lke}}

\subsection{ACT cluster likelihood}
\begin{figure}
\centering
\includegraphics[width=\columnwidth]{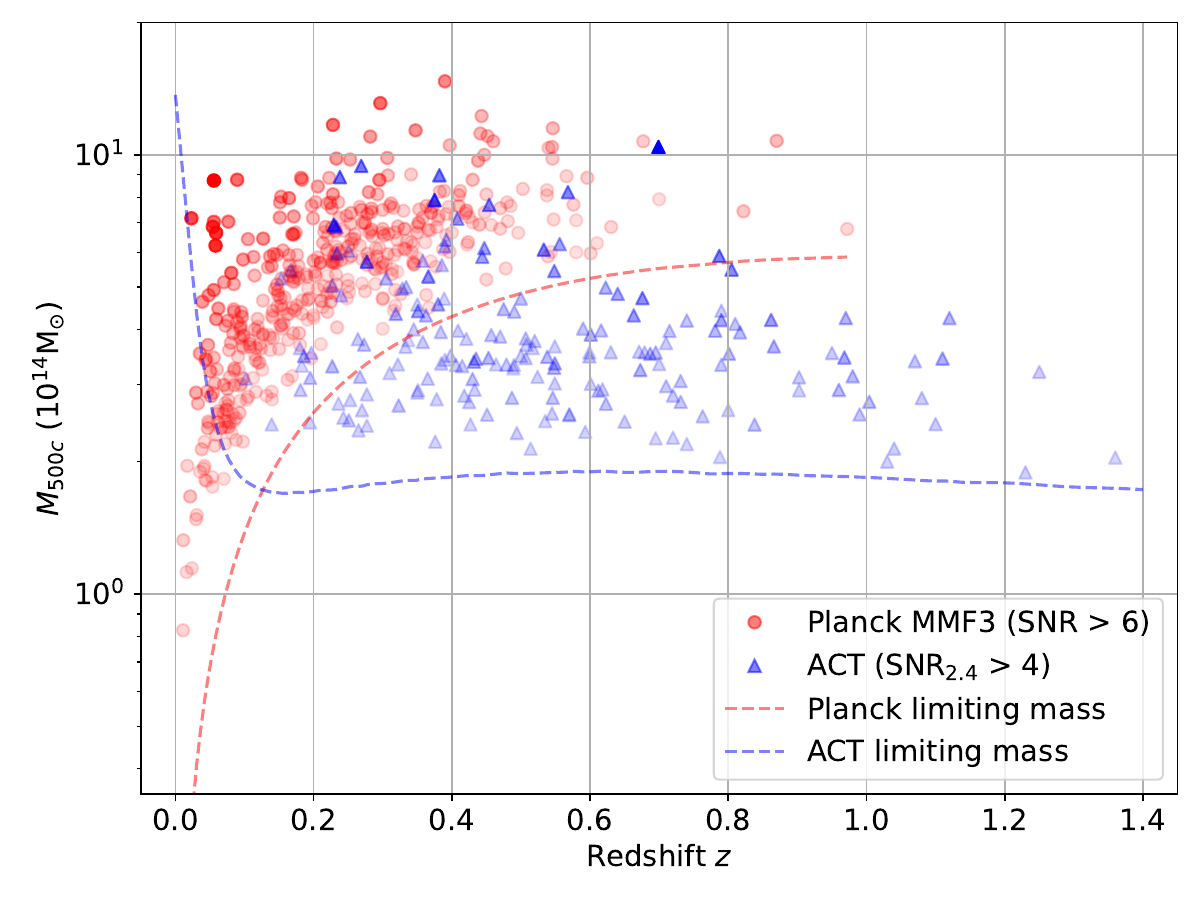}
\caption{Comparison of cluster sample of {\it Planck} (red circle) and ACT (blue triangle) in a redshift-mass plane. A sample with a higher signal-to-noise is shown as a darker colour. High confidence sample defined by SNR$_{2.4}$ cut of 5.6 for ACT contains 59 clusters (entire sample contains 182 clusters) and for the {\it Planck} the entire MMF3 sample is used as a cosmological sample (439 clusters). The dashed lines represent the limiting mass of the cluster sample which is calculated by the Compton-y parameter cut from each catalogue using $Y_{500}$ for {\it Planck} and $\widetilde{y}_0$ for ACT assuming a flat $\Lambda$CDM model with $\Omega_{\rm m} = 0.3$, $\Omega_{\Lambda} = 0.7$ and $H_{0} = 70$ km s$^{-1}$Mpc$^{-1}$ and all scaling parameter fixed to the fiducial values. SNR$_{2.4}$ refers to the SNR measured at a single scale of 2.4 arcmin.}
\label{fig:Mz_scatter}
\end{figure}
\begin{figure*}
\centering
\includegraphics[width=\textwidth]{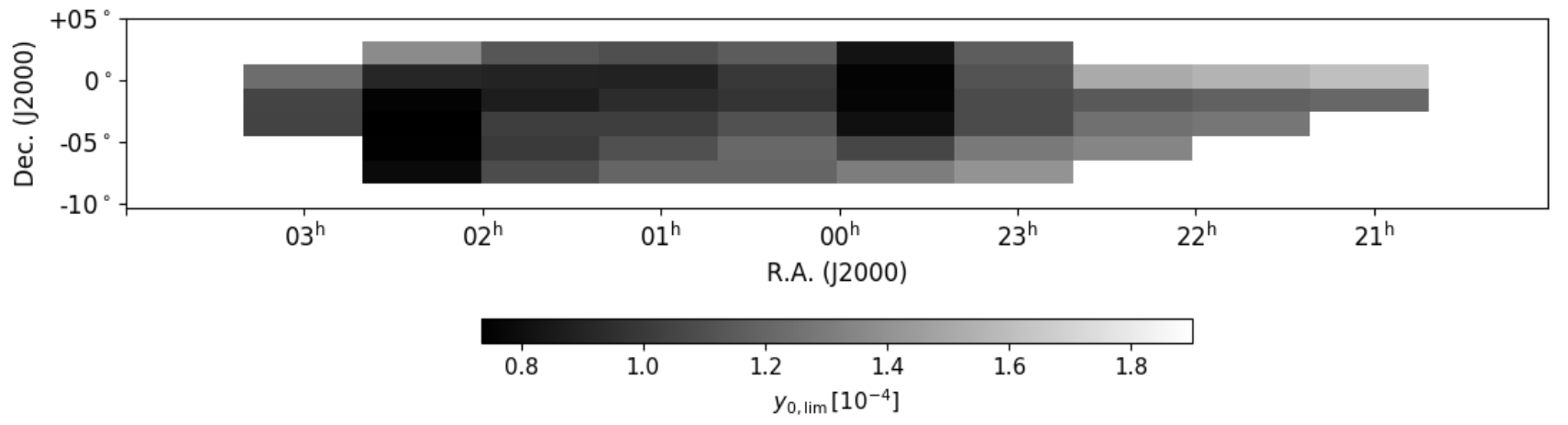}
\caption{The pixelated image of the limiting $\widetilde{y}_{0}$ map corresponding to SNR$_{2.4} = 5$ in Figure 9 of Hi18. Colour in grey scale represents the value of $\widetilde{y}_{0, {\rm lim}}$ in unit of $10^{-4}$. The number of equal-sized pixels in this map is 48 which appears to give a smooth distribution. The calculated number counts from the pixelated map image in a much smaller pixel size (227 pixels) gives the same number of clusters to within 3\%.}
\label{fig:act_ylim_map}
\end{figure*}
The ACT is a 6~m telescope located in the Atacama Desert in northern Chile.  The first SZ-detected cluster measurements from the Millimeter Bolometric Array Camera (MBAC) were released in \citealp{Marriage2011} and later 91 clusters were confirmed with redshifts in \citealp{Hasself2013}. With a polarization-sensitive receiver ACTPol (\citealp{Thornton2016}), a new SZ cluster catalogue of 182 clusters above SNR of 4 was published in \citealp{Hilton2017} (hereafter, Hi18). 
Hi18 provides two SNR measurements; the optimal SNR is defined as one which is maximised over all filter scales and SNR$_{2.4}$ is measured at a single filter scale of 2.4 arcmins. The cosmological sample of Hi18 is determined by SNR$_{2.4}$ cut of 5.6, which leaves 59 clusters. The ACT clusters were detected in the maps only at 148 GHz which uses spatial information rather than spectral characteristics of thermal SZ effect to select clusters in contrast to {\it Planck}. The survey area of Hi18 ($987.5\,\text{deg}^{2}$) covers both ACT Equatorial maps from MBAC and the ACTPol D56 field, which is referred to as the E-D56 field. 

Figure~\ref{fig:Nz_hist} displays the comparison of cluster sample of {\it Planck} and ACT used in this work. The distribution of the cosmological sample as a function of redshift is shown as a red circle for {\it Planck} and a blue triangle for ACT.  The {\it Planck} clusters are typically found at low redshifts whereas the ACT sample is spread in a wider redshift range due to the angular resolution of each telescope. The best-fit models from running the baseline likelihood are shown as histograms. A detailed description of them is described in Section~\ref{ss:each_sz_bias}.

In what follows we will describe the likelihood used in \citealp{Hasself2013} (hereafter, Ha13). The observable used by ACT is the central (peak) SZ flux $\widetilde{y}_{0}$\footnote{Tilde on $y_0$ means that the quantity is not corrected yet as it is measured from a fixed angular scale filter without a relativistic correction (following Ha13 description).} which is extracted at the single filter scale (2.4 arcmin). To relate $\widetilde{y_0}$ and the mass, $M_{500}$, Ha13 used a scaling relation 
\begin{equation}\label{eq:act_y0}
	\widetilde{y}_{0}= 10^{A_{0}} \left[\frac{M_{500}}{3\times 10^{14}M_{\odot}}\right]^{1+B_{0}} [E(z)]^{2} \ Q(\theta_{500})\,,
\end{equation}
where $10^{A_{0}}$ is the normalization and $B_{0}$ is the mass dependence of the scaling relation. The specific values of these constant scaling parameters used in Ha13 are presented in Table~\ref{table:scaling_params}.

$Q[\theta_{500}(M_{500},z)]$ is the filter mismatch function describing the difference between the expected cluster size measured by the filter and the actual size of the cluster. This accounts for the spatial convolution of the filter, the beam, and the shape of the integrated pressure profile. It plays the same role as taking different noise levels as a function of the filter scale $\theta_{500}$ as was done in the {\it Planck} analysis. 

To start with we assume a very simple case where the noise in the sky is the same everywhere. The observed flux limit could be set to a certain constant value by looking at the histogram of the observed flux, and we could recover the limiting mass using the given mass-observable scaling relation in Eq.~\eqref{eq:act_y0}. For illustrative purposes, the limiting mass recovered from a given limiting SZ flux is shown as a blue dashed line for ACT and also red for {\it Planck} in Figure~\ref{fig:Mz_scatter}. In the case of {\it Planck} a limiting mass is a steep curve in a redshift range where most of {\it Planck} clusters are detected whereas it is almost a horizontal line for ACT. The distribution of mass estimates from {\it Planck} (red dot) and ACT (blue triangle) catalogues used in this work is shown as a function of redshift in the figure. 

To be more accurate, we used the image of the limiting $y_{0}$ map in Figure 9 of Hi18 which illustrates that noise in the sky is not uniform.  The scanned image is pixelated and is shown in figure~\ref{fig:act_ylim_map} which gives the value of the limit of $\textrm{y}_{0}$ at each position of the map. The completeness is calculated for each sky patch and summed over as in the {\it Planck} likelihood. Note that the survey selection function is constructed by only using a single size filter $\theta_{500} = 2.4$ arcmin. For consistency with the {\it Planck} analysis in Eq.~\eqref{eq:plc_1d_complt}, the ACT theoretical number counts, taking into account the survey completeness using a noise map, can be expressed as
\begin{equation}\label{eq:act_complt}
    \chi_{_\textrm{act}} (z, M_{500}, \underline{x}) = \int \!\! d \ln y_{0} \,\text{P}[\ln y_{0} | \ln \overline{y}_{0}(z, M_{500})] \, \chi_{_\textrm{erf,act}}(y_{0}, \underline{x})\,,
\end{equation}
where the intrinsic scatter $\sigma_{\ln y_{0}}$ between the observed flux and the mass estimates are given by a log-normal distribution for $y_{0}$ around its mean value $\overline{y}_{0}$ for given mass $M_{500}$ and redshift $z$. The mean value is computed from the scaling relation given in Eq.~\eqref{eq:act_y0}, and we have that 
\begin{equation}\label{eq:act_scatter}
    \text{P}(\ln y_{0}\,|\ln \overline{y}_{0}) = \frac{1}{\sqrt{2\pi}\sigma_{\ln y_{0}}} \exp \left[-\left( \frac{\ln y_{0} - \ln \overline{y}_{0} (z, M_{500})} {\sqrt{2}\sigma_{\ln y_{0}}} \right)^{2}\right]\,.
\end{equation}

The survey selection function is given by $y_{0, \text{lim}}$ as a function of position $\underline{x}$ from the scanned map in figure~\ref{fig:act_ylim_map}
\begin{equation}\label{eq:act_erf}
    \chi_{_\textrm{erf,act}} (y_{0}, \underline{x}) = \frac{1}{2} \left[1 + \text{erf} \left( \frac{y_{0} - y_{0, \text{lim}}(\underline{x})} {\sqrt{2}\sigma_{y_{0}}} \right) \right]\,,
\end{equation}
where $\sigma_{y_{0}} =y_{0,\textrm{lim}}/ q_{\textrm{cut}}$.
As the limiting $y_{0}$ map represents the values cut by the SNR threshold of 5, the noise of the map could be expressed as the limiting $y_{0}$ value divided by 5. For the case of {\it Planck}, noise measurement is a function of filter scale and position in the map, on the other hand for ACT, the impact of variation in filter scale is taken into account in the $Q$ function hence the limiting $y_{0}$ is a function of the only position of the cluster. Overall, this approach resembles the {\it Planck} one-dimensional likelihood used in P14 according to Eq. ~\eqref{eq:plc_1d_erf}.

\begin{table}
    \caption{Summary of SZ observable-mass scaling parameters.}
    \begin{threeparttable}
    \centering
    \begin{tabular}{c c c}
    \hline \hline
	Parameters & Planck & ACT  \\ 
	\hline 
    Normalisation & $\log Y_{*} = -0.19 \pm 0.02$ &  $10^{A_{0}} = 4.95 \times 10^{-5}$ \\
    Mass dependence & $\alpha = 1.79 \pm 0.08$ & $B_{0} = 0.08$\\
    Redshift evolution & $\beta = 0.66$ &  \\
    Intrinsic scatter & $\sigma_{\ln Y} = 0.173 \pm 0.023$\tnote{a} & $\sigma_{\text{int}} = 0.2$ \\
    Reference & P14, P16 & Ha13, Hi18 \\
    \hline
    \end{tabular}
        \textbf{Note:} {\it Planck} scaling parameters are given with the central values and their error budgets. In the baseline analysis of P16 (two-dimensional likelihood), $\log Y_{*}$ and $\sigma_{\ln Y}$ are constrained by the Gaussian priors in the table, $\beta$ is fixed to the central value, and $\alpha$ is set free. The scaling parameters are derived using a process explained in detail in Appendix A in P14. ACT scaling parameter values were not varied in the cosmological analysis of Ha13 not for cluster characterisation in Hi18. \tnote{(a)} P16 used logarithmic intrinsic scatter value in computation instead of the natural logarithm. This is correctly included in this work.
    \end{threeparttable}
    \label{table:scaling_params}
\end{table}

\subsection{Connecting scaling relations and a universal likelihood} \label{sec:connect}

The SZ observable used by {\it Planck}, $Y_{500}$, is the integrated Compton-y signal within the angular size of $\theta_{500}$ and is used to estimate the cluster mass following $Y_{500}-M$ scaling relation in Eq~\eqref{eq:plc_Y500}. For ACT, the corresponding observable is the peak Compton-y signal, $\widetilde{y}_0$, which is translated into the cluster mass by Eq~\eqref{eq:act_y0}. In principle, these quantities are related and in fact, it is just a choice that is output by the cluster detection algorithm. In this section, we make connections between the two approaches in order to facilitate a joint likelihood analysis.

The key assumption in doing this is to suppose that a universal scaling relation can be applied to all the observed clusters. In other words, there is a single scaling relation that can be used in the likelihood for {\it Planck} and ACT data. It is clear from Eqs~\eqref{eq:plc_Y500} and~\eqref{eq:act_y0} that $Y_{*}$ and $10^{A_{0}}$ represent the normalization, while $\alpha$ and $B_{0}$ describe the power-law dependence of mass for the two scaling relations, respectively\footnote{The $Q$ function accounting for a mismatch between the filter scale and true cluster scale can be ignored when making these connections.}.

Using the small angle approximation, the integrated Compton-$y$ signal, $Y_{500}$, can be written as 
\begin{equation}\label{eq:connect_y}
	{Y}_{500} = \int_{0}^{\theta_{500}} 2\pi\theta \,y(\theta) \,{\rm d}\theta,
\end{equation}
where $y(\theta) = y_{0}\,f(\theta)$ is the product of a normalization factor $y_{0}$ and the pressure profile of cluster $f(\theta)$. If we now make the approximation $y(\theta)$ is constant, Eq.~\eqref{eq:connect_y} can be approximate as ${Y}_{500} \propto y_{0}\,{\theta_{500}}^2$.

Using the definitions of $Y_{500}$ in Eq~\eqref{eq:plc_Y500}, $\theta_{500}$ in Eq~\eqref{eq:plc_theta500}, and $\widetilde{y}_0$ in Eq~\eqref{eq:act_y0}, the coefficients related to the mass slope and the redshift evolution can be rather easily compared. In particular, we can find the mass variation exponents $\alpha$ and $B_0$ are given by 
\begin{equation}
    \alpha = \frac{5}{3} + B_0\,,
\end{equation}
while we can also read off $\beta=2/3$.

The Gaussian prior of the mass exponent used in P16 is $\alpha = 1.79 \pm 0.08$. With $B_{0} = 0.08$, which is used in Hi18, we obtain $\alpha (B_0) = 1.75$, which is within the error budget of the P16 prior. Conversely, the mass dependence parameter of Eq.~\eqref{eq:act_y0} can be derived from $\alpha$ as $B_{0}(\alpha) = 0.123$  which is slightly higher than the value used in Hi18. The value of the redshift exponent $\beta$ in the P16 baseline is fixed at $0.66$ which is the same as the corresponding value used in Hi18. 

In order to connect normalization factors between Eqs~\eqref{eq:plc_Y500} and~\eqref{eq:act_y0}, we need to calculate an integral using a pressure profile in Eq~\eqref{eq:connect_y}. This can be done by assuming that the gas in ICM is modeled following the Universal Pressure Profile (hereafter, UPP), which was used to build an average scaled profile of the given X-ray cluster sample at low redshift. The pressure profile is found to be approximately self-similar across a wide range of cluster masses and redshifts (see A10 for a detailed description). This is based on the generalised NFW (gNFW) model of \citealp{nagai2007} where
\begin{equation}\label{eq:dimensionless_P}
    \mathbb{P}(x) = \frac{P_{0}}{(c_{500}x)^{\gamma}[1+(c_{500}x)^{\alpha}]^{(\beta - \gamma)/\alpha}}\,,
\end{equation}
and $x=r/R_{500}$ with a physical radius $r$.
The scaling parameters for a pressure profile, $P_0$, $c_{500}$, $\alpha$, $\beta$, $\gamma$, are obtained from the best fitting model, which is summarised in Table~\ref{table:UPP_params}. 
\begin{table}
    \caption{gNFW profile parameters used in the UPP from A10. ACT scaling parameters are directly derived from the pressure model. {\it Planck} uses the UPP to construct a cluster model for cluster detection, but scaling parameters are fitted from observations. The values of profile parameters are taken from Eq.(12) in A10.}
    \begin{threeparttable}
    \centering
    \begin{tabular}{c c}
    \hline \hline
	Parameters & Values \\ 
	\hline
    P$_{0}$ & 8.403  \\ 
    c$_{500}$ & 1.177 \\ 
    $\alpha$ & 1.0510 \\
    $\beta$ & 5.4905 \\
    $\gamma$ & 0.3081 \\
    \hline
    \end{tabular}
    \end{threeparttable}
    \label{table:UPP_params}
\end{table}
With a dimensionless pressure profile defined by Eq.~\eqref{eq:dimensionless_P}, the physical pressure profile as a function of mass and redshift can be constructed,
\begin{equation}\label{eq:Pr}
     P(r) = P_{500} \left(\frac{M_{500}}{3\times 10^{14}M_{\odot}}\right)^{\alpha_{\mathrm{p}}(x)} \mathbb{P}(x)\,,
\end{equation}
where a characteristic pressure $P_{500}$ is given by
\begin{equation}\label{eq:p500}
    P_{500} = 1.65 \times 10^{-3} \text{keV}\,\text{cm}^{-3}\,[E(z)]^{\frac{8}{3}} \left(\frac{M_{500}}{3\times 10^{14}M_{\odot}}\right)^{\frac{2}{3}}\,,  
\end{equation}
and $\alpha_{\mathrm{p}}(x) = \alpha_{\mathrm{p}} + \alpha'_{\mathrm{p}}(x)$ represents the variation of normalisation of pressure profile depending on its mass.  Here, we have assumed $h=0.7$ for simplicity.
A10 finds a best-fitting value of $\alpha_{\mathrm{p}} = 0.12$, which is essentially a modification of the standard self-similarity, and this consequently gives a steeper mass dependence.  
The violation of self-similarity is described by
\begin{equation}
    \alpha'_{\mathrm{p}}(x) = 0.10 - (\alpha_{\mathrm{p}} + 0.10) \frac{(x/0.5)^3}{1 + (x/0.5)^3}\,,
\end{equation}
which leads to a change in the shape of the pressure profile as a function of mass. 

The pressure model used for cluster detection in \citealp{planck2014source} assumes that this mass dependence is almost constant with radius at $\Delta=500$ so it can be approximated to $\alpha'_{\mathrm{p}}(x) = 0$.
On the other hand, Ha13 keeps the additional mass dependence term in the pressure model, which can be rewritten as 
\begin{equation}
    \alpha_{\mathrm{p}}(x) = \alpha_{\mathrm{p}} + \alpha'_{\mathrm{p}}(x) = \frac{0.22}{1+8x^3}.
\end{equation}
Even though the additional mass dependence in the pressure profile is included and is parameterised as the mass dependence parameter $B_0$ in scaling relation in Ha13, when the normalisation $10^{A_0}$ in scaling relation is derived analytically, we ignore this additional mass dependence and hence within the analytical model for the pressure profile used to derive the normalisation of both integrated and peak signals the dependence is effectively the same.

Rewriting Eq.~\eqref{eq:connect_y} with a pressure profile in Eq.~\eqref{eq:Pr} with $\alpha_{\mathrm{p}}'(x)=0$ leads to
\begin{equation}
    {Y}_{500} = 2\pi y_{0} \left[\int_{-\infty}^{\infty} \mathbb{P}(t)\ {\rm d}t\right]^{\!-1} \int_{0}^{\theta_{500}} \int_{-\infty}^{\infty} \mathbb{P}\left(\sqrt{t^{2} + \left(\frac{\theta}{\theta_{500}}\right)^{2}}\right)\ {\rm d}t \,\theta {\rm d}\theta.
\end{equation}
Assuming $s = (\theta/\theta_{500})^2$ and $\theta_{500}^2 {\rm d}s = 2\theta {\rm d}\theta$, the integral is reduced to 
\begin{equation}\label{eq:y500_reduced}
    Y_{500} = X_{0} \,\pi y_{0} \theta_{500}^2
\end{equation}
where $X_0 = 0.1346$. 
Importantly, $X_0$ only depends on the slopes of the pressure profile (e.g., $\alpha,\beta,\gamma$), and on the UPP parameterization, it does not depend on cosmology.
Again, by plugging the definition of $Y_{500}$ in Eq.~\eqref{eq:plc_Y500}, $\theta_{500}$ in Eq.~\eqref{eq:plc_theta500}, and $y_0$ in Eq~\eqref{eq:act_y0} into Eq.~\eqref{eq:y500_reduced} and also setting the mass bias $1-b$ to be unity, we can find the relation between two normalization coefficients
\footnote{Note that there is a factor of 0.00472724 that multiplies in the Planck scaling relation, which comes from the pivot choices for the angular diameter distance and SZ flux normalisation, and conversion between radian to arcmin. All things combined yield $500^{-2}10^{-4}(\pi/180/60)^{-2}$.}
\begin{equation}
    Y_{*} = (4.378\times 10^{3}) \times 10^{A_{0}} \times 2^{\, \frac{5}{3} + B_{0}}.
\end{equation}

Using parameter values from Ha13, we find that the calculated {\it Planck} normalization is $Y_{*}(10^{A_{0}}, B_{0}) = 0.727$ whereas that used in P16 is $Y_{*} = 0.652 \pm 0.03$. Conversely, we could derive the normalisation factor for the ACT scaling relation from {\it Planck} scaling parameters, which gives $10^{A_{0}}(Y_{*},\alpha) = 4.30 \times 10^{-5}$. 
It is within the error budget of less than 3$\sigma$. 

Overall we conclude that, while the models used in P14/P16 and Ha13/Hi18 are not identical, they are closely related and their normalization is similar. In section~\ref{ss:combined} we will use the P14/P16 model as a universal cluster scaling relation. In this sub-section we have derived relations that will allow us to calculate a scaling relation for $y_0$ from that for $Y_{500}$ and $\theta_{500}$ assuming the UPP from A10.

\section{External data used in this paper} \label{sec:external}
Cluster abundance is a powerful tool to constrain the cosmic matter density $\Omega_{\rm m}$ and the amplitude of density fluctuation at present $\sigma_8$ if we have an accurate normalization of the scaling relation or precise knowledge of the mass bias.  It is also sensitive to some other cosmological parameters such as the Hubble parameter, $H_0$, and the dark energy equation of state parameter, $w$. In order to extract estimates for these parameters, it is necessary to combine information from other sources.  In what follows we will use data from a number of other probes and these are detailed in this section.

\subsection{Weak lensing based mass bias estimates}\label{ss:weak_lensing_priors}

 Weak lensing mass measurements of clusters present a potentially powerful way of measuring the mass of clusters. In principle, it is the most direct way of measuring the total mass without assuming a dynamical state of a cluster, for example, hydrodynamic equilibrium or spherical symmetry of gas. However, for a variety of reasons, it is only possible for a small number of clusters within the sample. Therefore, the approach that has become standard is to parameterize the difference between the assumed true mass obtained from weak lensing measurements and that deduced from X-ray observations using the assumption hydrostatic equilibrium using the mass bias parameter, $1-b$.  Here, we will make the strong assumption that it is a weak function of mass and redshift, largely because there is not enough information available to do much more.

P16 used three different weak lensing mass priors on $1-b$. The baseline was from the Canadian Cluster Comparison Project (CCCP, \citealp{Hoekstra2015}) which is given as $1-b = 0.780 \pm 0.092$.  In addition, they used  $1-b = 0.688 \pm 0.072$ from Weighing the Giants (WtG, \citealp{vonderL2014}) and CMB lensing which is given as $1-b = 1.01_{-0.16}^{+0.24}$ (CMBlens, \citealp{Melin2015}). CCCP and WtG are obtained from shear mass measurements of a subset of the P16 clusters using the shapes of background galaxies\footnote{\citealp{Battaglia_2016} showed that Eddington bias correction shifts a mass bias estimate of CCCP (\citealp{Hoekstra2015}) and WtG to a lower side.}. CMBlens also uses gravitational weak lensing but from the effect on the CMB temperature anisotropies. 

Here, we use the two updated priors, one from galaxy shear and another from CMB. CCCP has been combined with Multi Epoch Nearby Cluster Survey (MENeaCS) using 61 {\it Planck} clusters in \citealp{Herbonnet2019cccp} to yield a new constraint, $1-b = 0.84 \pm 0.04$, while \citealp{Zubeldia2019} used the CMBlens mass measurement technique for 433 Planck MMF clusters to yield $1-b = 0.71 \pm 0.10$. In both cases, uncertainties have been reduced significantly compared to the ones from priors used in P16, and this will lead to reductions in the uncertainties in the cosmological parameters. 

We note that there are other measurements\footnote{The Local Cluster Substructure Survey (LoCuSS, \citealp{Smith_2015}) finds $1-b = 0.95 \pm 0.04$ using 44 Planck clusters. This result is not compatible with the values found by other measurements and indeed is in contradiction with results from a hydrodynamic simulation of cluster formation and, therefore, we have not used it in our analyses.}. The Cluster Lensing And Supernova survey with Hubble (CLASH, \citealp{PennaLima2017}) measures $1-b = 0.73 \pm 0.10$ using 21 {\it Planck} clusters, whereas 8 ACT clusters have been observed using the Subaru Hyper Supreme-Cam (HSC, \citealp{Miyatake_2019}) and the mass bias is found to be $1=b=0.74_{\small -0.12}^ {\small +0.13}$.  In order to make it easier to present our results we have not used these priors, and indeed also WtG from P16. Suffice to say that they will yield similar results to the CMBlens measurement from \citealp{Zubeldia2019}.  Information of all recent weak lensing mass measurements is summarized in Table~\ref{table:priors} with the two priors that we will use in bold. 

\begin{table}
    \centering
    \caption{Mass bias priors with 1$\sigma$ uncertainty from recent mass calibrations described in Section~\ref{ss:weak_lensing_priors}. The first three lines of the table show the priors used in P16. In the middle, we show the recent weak lensing measurements from galaxies using 21 {\it Planck} clusters (CLASH) and 8 ACT clusters (HSC) which we do not use, but we note that they are similar to the CMBlens prior at the we do use. The updated CCCP and CMBlens mass bias priors in the final two lines of the table are those that are used in the paper and they are denoted CCCP and CMBlens respectively in the subsequent sections.} 
    \begin{tabular}{l c c} 
	\hline \hline
	Prior name & Value and Gaussian errors & Reference\\ \hline
 	WtG (P16) & $0.688 \pm 0.072$ & \citealp{vonderL2014} \\
 	CCCP (P16) & $0.780 \pm 0.092$ & \citealp{Hoekstra2015}\\
 	CMBlens (P16) & $1.01^{\small +0.24}_{\small -0.16}$ & \citealp{Melin2015} \\
    \hline 
 	CLASH & $0.73 \pm 0.10$ & \citealp{PennaLima2017} \\
 	HSC & $0.74_{-0.12}^{+0.13}$ & \citealp{Miyatake_2019} \\ 
 	\hline
 	{\bf CCCP}  & $0.84 \pm 0.04$ & \citealp{Herbonnet2019cccp}\\ 
 	{\bf CMBlens} & $0.71 \pm 0.10$ & \citealp{Zubeldia2019} \\ \hline
    \end{tabular}
    \label{table:priors}
\end{table}

\subsection{BAOs}

The Baryonic Acoustic Oscillations (BAO) are sensitive to the cosmic matter density, $\Omega_{\rm m}$, and Hubble parameter, $H_0$, via the line-of-sight comoving distance and angular diameter distance. Since the dependence of the clusters counts on $H_0$ is relatively weak, BAO data is commonly used when primary CMB is not used to remove this degeneracy. We use the latest data from The Sloan Digital Sky Survey (SDSS) IV Extended Baryon Oscillation Spectroscopic Survey (eBOSS) \citep{Alam_2021} which contains SDSS DR7 Main Galaxy Sample (MGS) \citep{Ross_2015, Howlett_2015}, BOSS DR12 Galaxies \citep{Alam_2017}, eBOSS Luminous Red Galaxies (LRGs) \citep{Bautista_2020}, eBOSS Emission Line Galaxies (ELGs) \citep{de_Mattia_2020}, eBOSS Quasars \citep{Neveux_2020}, and SDSS Lyman-$\alpha$ Forest \citep{du_Mas_des_Bourboux_2020}. MGS, BOSS galaxies, LGS, and ELG data cover a low redshift range and the rest cover a relatively higher redshift range.  We note that this is an updated compilation of BAO measurements and is not the same as that used in P16. However, it is a very similar dataset used in \citealp{Planck2018vi} and is likely to lead to very similar effects measured cosmological parameters with strengthened constraining power.

\subsection{Big Bang Nucleosynthesis and spectral index priors}

Cluster abundances are also weakly sensitive to the physical baryon density parameter, $\Omega_{\rm b}h^2$, and the spectral index of density perturbations, $n_{\mathrm{s}}$. They can modify the shape of the power spectrum on the scales relevant to cluster formation.  In order to control these parameters, but not fix them, in our analyses when we do not include primary CMB data, we will always use a prior from Big Bang Nucleosynthesis (BBN) $\Omega_{\rm b}h^2=0.022\pm 0.002$ \citep{steigman2008}.  In addition, we adopt a prior on $n_{\rm s}$ from \citealp{Planck2018vi} ($n_{\rm s} = 0.9648 \pm 0.0042$). 
For brevity, the combined action of these two priors will be denoted BBN in the subsequent presentation of the results.  
 
\subsection{CMB} \label{sec:cmb}
 
The power spectrum from primary CMB temperature anisotropies provides measurements of baryon and cold dark matter densities, $\Omega_{\rm b}$ and $\Omega_{\rm c}$, as well as the amplitude of the initial power spectrum of the fluctuations, $A_{\rm s}$ and is used in some of the analyses presented in this paper. In particular, the CMB data constrains the acoustic angular scale, $\theta_{\rm s}$, to very high precision. The {\it Planck} primary CMB data (TT,TE,EE+lowP) used in P16 from \citealp{Planck2015xiii} is referred to as PCMB16. We use the most recent CMB data (TT,TE,EE+lowE+lensing) from \citealp{Planck2018vi} (hereafter, PCMB20) in this work.

\section{Comparison of Planck and ACT likelihoods}\label{ss:result_each}

\begin{table}
    \centering
    \caption{Nuisance and cosmological parameters of the baseline dataset for likelihood computation, following the definitions in P16.}
    \begin{tabular}{c|c|c|c} 
        \hline \hline
        Parameters & Fixed & Varied & Gaussian priors \\ \hline 
         $\log Y_*$ & & $[-0.291, -0.081]$ & $-0.186 \pm 0.021$\\
         $\alpha$ & & [1, 3] &  \\
         $\beta$ & 0.66 & & \\
         $\sigma_{\ln Y}$ & & [0.1, 0.3] & $0.173 \pm 0.023$ \\ 
         $1-b$ & & [0.7, 1.0] & $0.84 \pm 0.04$ \\ \hline
         $H_{0}$ & & [55, 90] & \\
         $n_{\mathrm{s}}$ & & [0.8, 1.2] & $0.9649 \pm 0.0042$ \\
         $\Omega_{\mathrm{b}} h^2$ & & [0.012, 0.032] & $0.022 \pm 0.002$ \\ 
         $\Omega_{\mathrm{c}} h^2$ & & [0.05, 0.19] & \\ 
         $\ln (10^{10} A_{\mathrm{s}})$ & & [2, 4] & \\
         $100\theta_{\mathrm{MC}}$ & & [0.5, 10] & \\
         $\tau$ & 0.09 & & \\
         $\sum m_{\nu}$ & 0.06 & & \\
         $N_{\mathrm{eff}}$ & 3.046 & & \\ \hline
    \end{tabular}
    \label{tab:params_all}
\end{table}
    
In this section, we reproduce {\it Planck} SZ cluster counts analyses from P16 using a version of the likelihood which we have incorporated into {\tt COBAYA}\footnote{\url{https://github.com/CobayaSampler/cobaya}} \citep{cobaya2021} with updated prior information. 
In addition, we have performed an analysis using a binned likelihood with the ACT cluster catalogue from Hi18. A full dataset\footnote{\url{https://lambda.gsfc.nasa.gov/}} and a code used ({\tt CosmoMC}\footnote{\url{https://github.com/cmbant/CosmoMC}}, \citealp{Lewis2002}) for P16 are publicly available, and we have found good agreement with it, but no publicly available cluster likelihood for Hi18.

The baseline will be the two-dimensional likelihood over the distribution of redshift and SNR. Following P16, the {\it Planck} sample is split into 10 redshift bins $(\Delta z = 0.1)$ and 5 SNR bins $(\Delta \log q = 0.25)$.
For ACT, the Hi18 sample is split into 12 redshift bins with the same bin size used in P16 but only two bins for SNR.  Assuming the flat $\Lambda$CDM model, we vary the physical baryon and cold dark matter density parameters $\Omega_{\rm b}h^2$ and $\Omega_{\rm c}h^2$, $\theta_{\rm MC}$, which is the approximation to the acoustic angular scale $\theta_{\rm s}$, the normalisation and spectral index of matter power spectrum $A_{\rm s}$ and $n_{\rm s}$ to derive the total mass density parameter $\Omega_{\rm m}$, amplitude of matter power spectrum fluctuation $\sigma_8$, and the Hubble parameter $H_0$.  The optical depth of reionisation $\tau$ is fixed unless the CMB data is included in the likelihood. The baseline includes BAO data and BBN priors on $\Omega_{\rm b}h^2$ and $n_{\rm s}$. The nuisance parameters follow the P16 baseline setting; the normalisation $Y_*$ and the intrinsic scatter $\sigma_{\ln Y}$ are varied with Gaussian priors on them, the mass dependence $\alpha$ is varied free, and the redshift dependence $\beta$ is fixed to its standard value unless stated otherwise. The settings for cosmological and nuisance parameters are summarised in Table~\ref{tab:params_all}.

\subsection{Pipeline validation: Planck likelihood}\label{ss:planck_val}

\begin{figure}
\centering
\includegraphics[width=0.9\columnwidth]{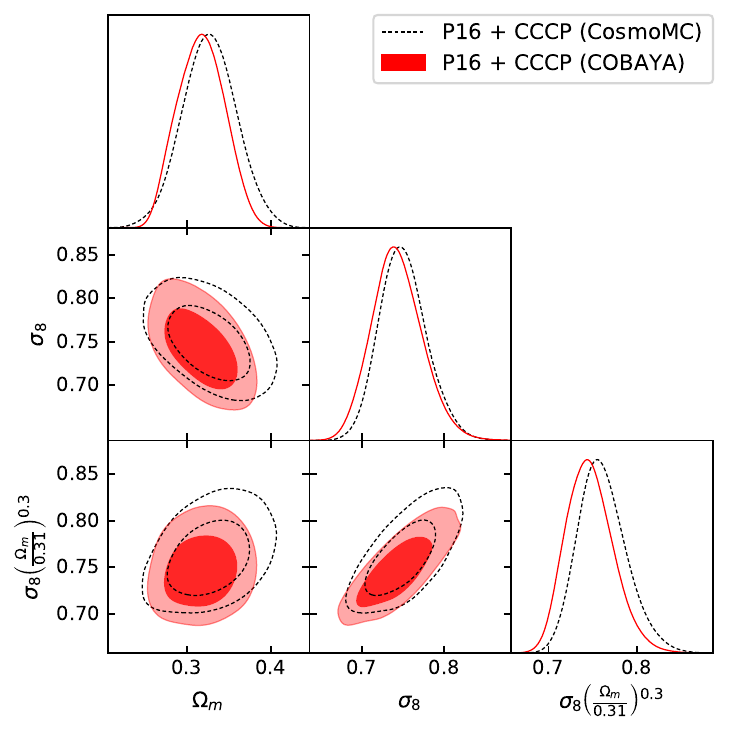}
\caption{We present a reproduction of the  P16 results using {\tt COBAYA}. P16 using {\tt CosmoMC} are shown as dashed black contours and results from {\tt COBAYA} are the red contours. {\tt CosmoMC} constraints are directly from the chains obtained from P16 analysis and plotted together with the constraints from {\tt COBAYA}. The dataset used is for the baseline, which is with BAO, BBN, and CCCP prior to mass bias. The details are described in  Table~\ref{tab:params_all}. The two likelihoods agree to a high degree with the differences in the means being less than half of the standard deviation in the marginalised 1D posteriors.}

\label{fig:planck2015result}
\end{figure}
%
As a pipeline validation, we reproduce P16 results using {\tt COBAYA} with the exact same dataset and priors and compare them to the results from {\tt CosmoMC}. {\tt CosmoMC} constraints are directly from the chains obtained from P16 analysis and plotted together with the constraints from {\tt COBAYA}. Note that the BAO and CCCP information used here are not updated for the purpose of comparison. Figure~\ref{fig:planck2015result} shows the cosmological constraints from the {\it Planck} dataset with BAO, BBN, and CCCP priors. 
Results from {\tt COBAYA} are shown in red-filled contours and corresponding constraints from {\tt CosmoMC} are shown in empty black contours with dashed lines. Cosmological constraints using {\tt COBAYA} are $\sigma_8 = 0.74 \pm 0.03$, $\Omega_{\rm m} = 0.32 \pm 0.03$, and $\sigma_{8} (\Omega_{\rm m}/0.31)^{0.3} = 0.747 \pm 0.027$ and using {\tt CosmoMC} are $\sigma_8 = 0.75 \pm 0.03$, $\Omega_{\rm m} = 0.33 \pm 0.03$, and $\sigma_{8} (\Omega_{\rm m}/0.31)^{0.3} = 0.761 \pm 0.028$. As can be seen in the figure, results are in good agreement overall with the differences in the means values being less than half of the standard deviation for the three parameters. 



\subsection{Pipeline validation: ACT likelihood}\label{ss:ACT2013}

\begin{figure}
\centering
\includegraphics[width=0.9\columnwidth]{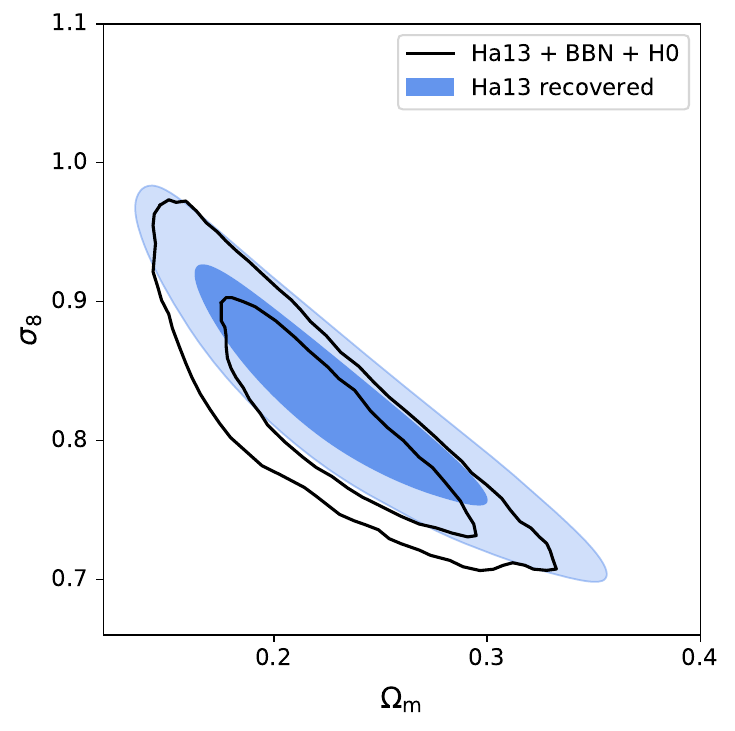}
\caption{Constraints on $\Lambda$CDM cosmology from ACT SZ + BBN + $H_0$ using out ACT likelihood in light blue contours. Over-plotted in black are the cosmological constraints from the top of Figure 14 in Ha13 for comparison. This illustrates a good level of agreement between our ACT likelihood with previous results.}
\label{fig:act2013result}
\end{figure}
%
As a check of our ACT likelihood, we have attempted to compare with the cosmological constraints from Ha13. The cosmological sample for this analysis is 15 clusters above the SNR threshold of 5.1 over the 270 square degrees survey area in the ACT Equatorial map. Ha13 compares the results from different pressure profile models and also adds the dynamical mass measurements from galaxy velocity dispersion. Their baseline uses the UPP model with fixed scaling parameters and same values for scaling parameters $10^{A_{0}}$, $B_{0}$ and the intrinsic scatter $\sigma_{\rm int}$ are used for cluster characterisation in their later work (Hi18), but there is an additional scaling parameter $C_{0}=-0.025$ in Ha13 that accounts for an additional mass dependence in the mismatch function $Q(\theta_{500}/m^{C_{0}})$ where $m \equiv M_{500}/(3\times 10^{14}M_{\odot})$. In addition, they modelled a corrects for the relativistic SZ effect which we also included - although we have not included it in the rest of our analyses. The effective centre of the frequency band is shifted slightly (146.9 GHz, \citealp{Swetz2011}), and this us is used to calculate relativistic correction. Not all the information needed for the completeness computation is readily available, so first we digitised the Q function as a function of a filter scale from Figure 6 in Ha13 and we also scanned and digitised the image of the filtered noise map at 148 GHz from Figure 1 in Ha13. The map is filtered with a fixed scale of $\theta_{500} = 5.9$ arcmin to detect clusters and provides the sensitivity to the cluster detection process. 

The Ha13 is based 15 ACT clusters with priors of $\Omega_{b}h^{2} = 0.022 \pm 0.002$ and $H_{0} = 73.9 \pm 3.6$ km s$^{-1}$Mpc$^{-1}$. Scaling parameters were fixed to the fiducial values assuming the UPP model and the mass bias parameter was not included. Figure ~\ref{fig:act2013result} shows the comparison of cosmological constraints from Ha13 with the results from the pipeline we use for ACT analysis.  Since the values of cosmological constraints for this dataset are not given in Ha13, the corresponding contour is scanned and over-plotted from Figure 14 in Ha13 for comparison (black contour). The filled blue contour is the result for the same Ha13 clusters using our likelihood. We find $\sigma_8 = 0.833 \pm 0.058$ and $\Omega_{\rm m} = 0.229 \pm 0.045$ for Ha13 clusters.

\subsection{Results from SZ datasets using mass bias priors} \label{ss:each_sz_bias}
\begin{figure}
\centering
\includegraphics[width=\columnwidth]{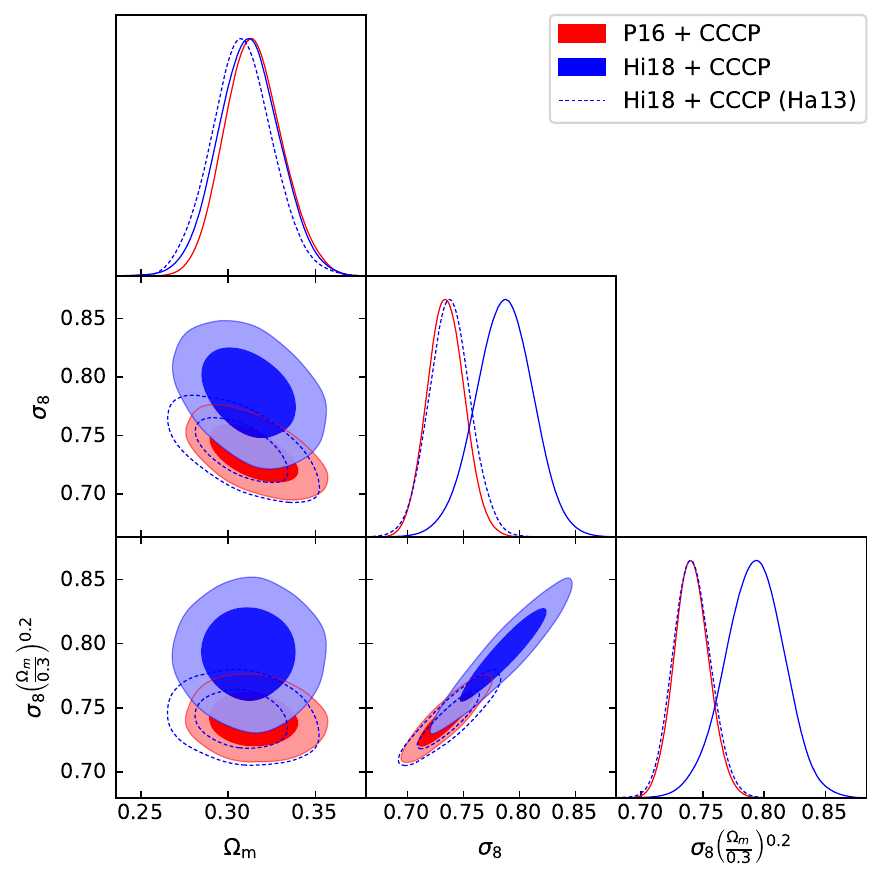}
\caption{Comparison of the constraints on cosmological parameters from SZ cluster data of P16 (red) and Hi18 (blue) with the CCCP weak lensing measurement prior to mass bias. As the CMB information is not included, we adopt the BAO data and a BBN prior on $\Omega_{\rm b}h^{2}$ and $n_{\rm s}$. The unfilled blue contours with the dashed lines are the constraints where ACT scaling parameters are fixed to the fiducial values used in Ha13. The blue-filled contours are the constraints where {\it Planck} cluster model is used for the Hi18 clusters. Note that the discrepancy in $\sigma_8$ constraints is likely caused by the fact that we choose the same mass bias prior, which is effectively the normalisation of the scaling relation.}
\label{fig:sz_cccp_each}
\end{figure}
%
%
%
\begin{table*}
    \centering
    \begin{tabular}{c c c c c}
    \hline \hline
     Parameters & P16 + CCCP & Hi18 + CCCP & P16 + CMBlens & Hi18 + CMBlens  \\ \hline
     $\Omega_{\rm m}$ & $0.315 \pm 0.017$ & $0.312 \pm 0.018$ & $0.317 \pm 0.016$ & $0.313 \pm 0.018$ \\
     $\sigma_{8}$ & $0.734 \pm 0.017$ & $0.786 \pm 0.026$ & $0.781 \pm 0.040$ & $0.839 \pm 0.047$ \\
     $\sigma_{8} (\Omega_{\rm m}/0.3)^{0.2}$ & $0.741 \pm 0.014$ & $0.792 \pm 0.024$ & $0.790 \pm 0.041$ & $0.846 \pm 0.048$ \\
     \hline
    \end{tabular}
    \caption{Cosmological constraints from individual cluster dataset using mass bias priors from CCCP and CMBlens - the updated values tabulated in Table~\ref{table:priors}. In all cases, the BAO data is used as well as the BBN prior.}
    \label{tab:sz_cccp_cmblensing}
\end{table*}
%
%
%
%
%
%
\begin{figure}
\centering
\includegraphics[width=0.9\columnwidth]{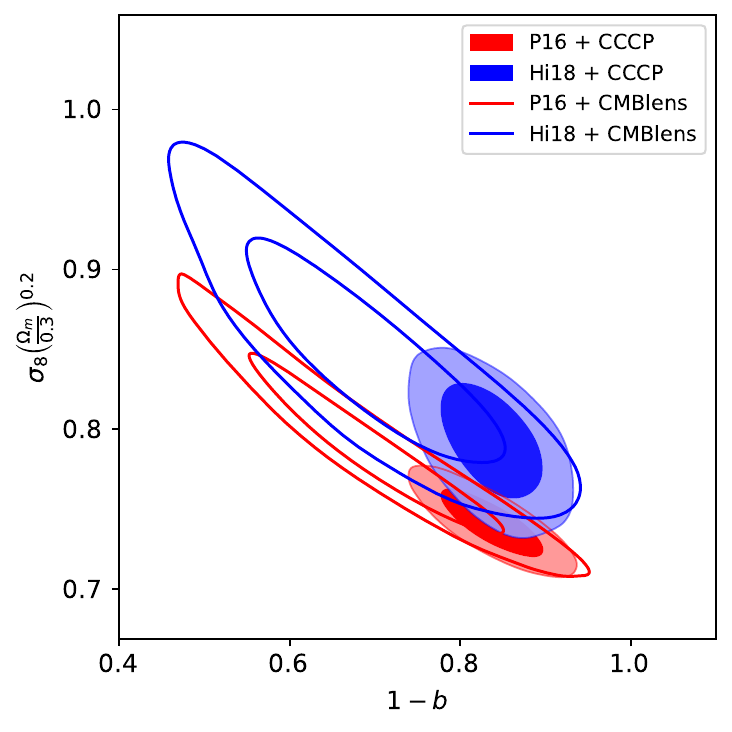}
\caption{Comparison of the constraints on cosmological parameter combination and mass bias from SZ cluster data of P16 (red) and Hi18 (blue) with the CCCP (filled) and CMBlens (unfilled) prior to mass bias. BAO data and BBN priors are included in the datasets for all cases. Constraints on $\sigma_{8} (\Omega_{\rm m}/0.3)^{0.2}$ is sensitive to the choice of mass bias priors.}
\label{fig:sz_cccp_cmblensing}
\end{figure}
%

We now investigate what one can deduce from the SZ datasets with the minimum amount of external information. As explained earlier, the baseline for this and our subsequent analysis will be to use the Planck cluster model for both datasets, that is, we will presume that there is a universal cluster population. However, here we have performed two analyses for Hi18: one using the fixed values of $A_0$, $B_0$ and $\sigma_{\rm int}$ in a similar way to Ha13, which we will call the ACT cluster model. The results of these using the CCCP and BBN priors are presented in Figure~\ref{fig:sz_cccp_each}. The values of constraints on cosmological parameters found in this section are summarised in Table~\ref{tab:sz_cccp_cmblensing}. 

Using  the ACT cluster model for the Hi18 data (unfilled blue), we find $\sigma_8 = 0.738 \pm 0.019$, $\Omega_{\rm m} = 0.308 \pm 0.018$, and $\sigma_{8} (\Omega_{\rm m}/0.3)^{0.2} = 0.741 \pm 0.015$. We see that the P16 contours agree rather well with those from Hi18 using the ACT cluster model, which appears to be a strange coincidence. 

When we use the Planck cluster model, we find the P16 and Hi18 contours agree less well - a  2.7$\sigma$ difference. For the Hi18 sample, the size of contours is increased as one might expect, due to the fact that the scaling parameters are no longer fixed, and a higher value of $\sigma_8$ constraint is obtained.
 
The choice of the CCCP prior to the mass bias is crucial to these conclusions and therefore it is sensible to compare to alternatives and we have chosen to use the updated CMBlens measurement from \citealp{Zubeldia2019}. 
Figure~\ref{fig:sz_cccp_cmblensing} shows constraints on cosmological parameter combination $\sigma_{8} (\Omega_{\rm m}/0.3)^{0.2}$ versus mass bias again using the BAO and BBN priors. Red contours represent P16 and blue ones represent Hi18 with the Planck cluster model. The filled contours are from using the CCCP prior and the unfilled contours are from using the CMBlens prior.  It is clear that, due to the well-known degeneracy between the mass bias and $\sigma_{8} (\Omega_{\rm m}/0.3)^{0.2}$ the cosmological constraints are very sensitive to which prior one uses. Constraints from different mass biases in Figure~\ref{fig:sz_cccp_cmblensing} are summarised in Table~\ref{tab:sz_cccp_cmblensing}.  These represent the state-of-the-art constraints from the individual experiments, {\it Planck} and ACT, on cosmological parameters.

\subsection{Results from combining SZ datasets with CMB data }
\begin{figure}
\centering
\includegraphics[width=\columnwidth]{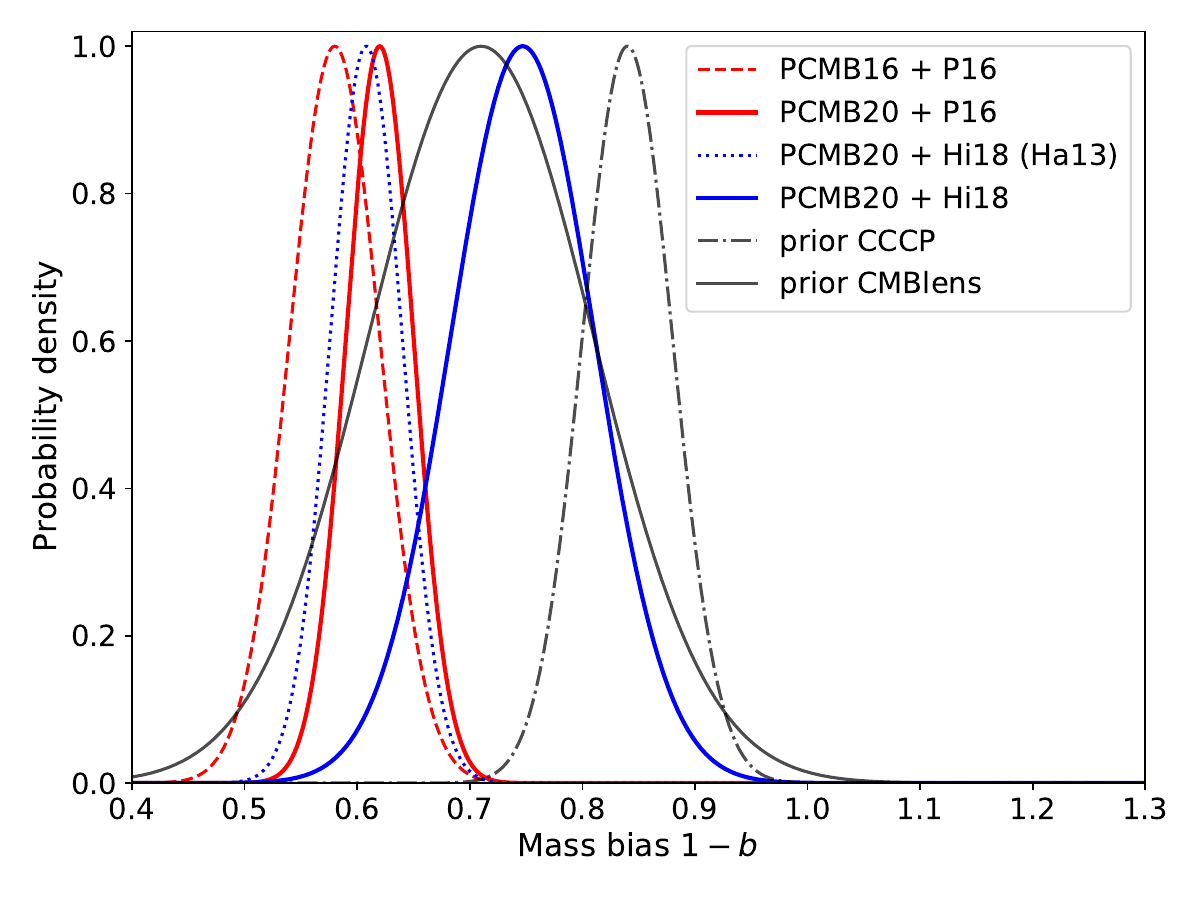}%
\caption{Comparison of the constraints on the mass bias from cluster data of P16 (red) and Hi18 (blue) experiments combined with the {\it Planck} CMB data. Constraints are obtained from combining P16 with PCMB16 (dashed red) and with PCMB20 (solid red) which gives $1-b=0.612 \pm 0.030$. The Hi18 cluster data combined with PCMB20 returns the mass bias value of $1-b = 0.608 \pm 0.032$ when using fixed fiducial ACT scaling parameters $10^{A_0}= 4.95\times 10^{-5} $ and $B_0=0.08$ from Ha13 (dotted blue) which is very close to the value obtained from P16. When the P16 cluster model is used for the Hi18 cluster sample, we find the mass bias constraint of $1-b = 0.747 \pm 0.064$ (solid blue). The updated weak lensing mass measurements used as priors in this work are also shown in the figure as grey lines; CMB lensing (solid, \protect\citealp{Zubeldia2019}), CCCP (dashdot, \protect\citealp{Herbonnet2019cccp}).}%
\label{fig:mass_bias_prob}
\end{figure}
%
In this section, we will combine the SZ likelihoods for P16 and Hi18 with the CMB data described in Section~\ref{sec:cmb} which will allow us to deduce the value of $1-b$ which best fits the SZ data in a cosmology defined by the CMB data. 
We will concentrate on using the most recent CMB data from PCMB20 in what follows.
We find that using this data $1-b=0.61 \pm 0.03$ compared to $1-b=0.62 \pm 0.03$ from PCMB20, which shows that our {\tt COBAYA} implementation faithfully reproduces the {\tt CosmoMC} version.

Figure~\ref{fig:mass_bias_prob} presents the mass bias constraints from the primary CMB from PCMB16 and an update from PCMB20 along with the equivalent for Hi18 combined with PCMB20. 
The weak lensing measurements imposed as priors on the mass bias parameter in the subsequent sections are included (grey lines) for comparison. 
The best-fit mass bias value for the Planck cluster counts using PCMB20 (solid red) is slightly larger than the value using PCMB16 (dashed red). 
This shift is caused by the best fitting value for $\tau$ in PCMB20 which is ultimately down to an improved analysis of polarization data presented in PCMB20. 
This value is still lower than the central values of updated weak lensing priors and appears to be at odds with the updated CCCP prior.

\begin{figure}
\centering
\includegraphics[width=0.9\columnwidth]{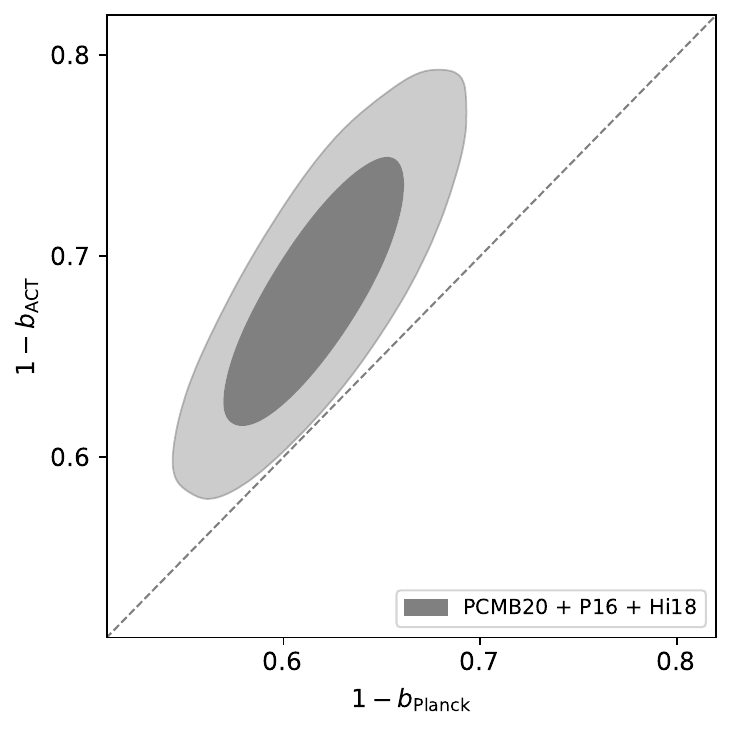}
\caption{Comparison of constraints on mass biases of P16 and Hi18 when combined with PCMB20. Scaling parameters are varied by {\it Planck} fiducial values but we keep the mass biases for P16 and Hi18 separately in the scaling relation. When two cluster datasets are combined, the mass bias for P16 is dragged to a slightly higher value and the mass bias for Hi18 is dragged to a slightly lower value.}
\label{fig:cmb_plcactSZ_twobiases}
\end{figure}
%

The best fitting value of mass bias when combining Hi18 and PCMB20 is $1-b=0.608 \pm 0.032$ using fixed values of $A_0$, $B_0$ and $\sigma_{\rm int}$ as done in Ha13 (dotted blue). 
This is surprisingly close to the value obtained from PCMB20+P16 and prima-facie appears to be a significant success. 
However, one is not comparing like with like. 
Fixing the scaling parameters leads to an artificially small uncertainty which is coincidentally similar to that from PCMB20+P16. 
Moreover, when the full {\it Planck} cluster model - making the assumption that there is a universal cluster model - is used the central value shifts and we find that the mass bias constraint is somewhat higher, but with a larger uncertainty $1-b = 0.747 \pm 0.064$ (solid blue). 
This is more in keeping with the values found in weak lensing analyses but is still compatible with the PCMB20+P16 results with the uncertainties at 95\% confidence, suggesting that there is a basic agreement between the two experiments. 
This compatibility motivates combining the two cluster datasets together as we do in Section~\ref{ss:combined} in order to probe the cluster model, or even the late-time evolution of the Universe, both of which can improve the correspondence between the datasets. 

The result of combining both SZ datasets with the PCMB20 data, but allowing different mass bias parameters for {\it Planck} and ACT is presented in Figure~\ref{fig:cmb_plcactSZ_twobiases}. The marginalised constraints on the individual mass biases are $1-b_{\rm plc}=0.617 \pm 0.031$ and $1-b_{\rm act}=0.684 \pm 0.044$ which are somewhat different from those.
However, the correlation illustrates that there is a slightly more complicated picture since the result illustrates that there is a systematically slightly higher bias found using the Hi18 data even though they are overlapping in their uncertainty ranges as shown in Figure~\ref{fig:cmb_plcactSZ_twobiases}. 
We will return to this point in Section~\ref{ss:combined}.
It is worth noting that the ACT mass bias constraint obtained here is consistent with the result, which is obtained independently, using the redMaPPer richness-based weak lensing mass calibration in Hi18 (see Section 6.1 in \citealp{Hilton2017} for a detailed description).


\section{Combining Planck and ACT cluster counts}
\label{ss:combined}

In Section~\ref{ss:result_each}, we showed that the constraints obtained from P16 and Hi18 are compatible with each other if one uses the same cluster model. 
In this section, we combine datasets and extract the constraints assuming the same cluster model and that the clusters have the same bias. 
The fact that they do not exactly agree and that they probe different parts of the cluster population, both in redshift and mass, (see Figure~\ref{fig:Mz_scatter}) suggests that such an analysis is likely to allow constraints on other parameters not previously thought possible using the individual datasets.
For the analysis of the Hi18 clusters, we will compute $10^{A_0}$ and $B_0$ from $Y_*$ and $\alpha$ using the relation derived in Section~\ref{sec:connect}. 
The normalisation parameter $Y_*$ is varied with a Gaussian prior of $\log Y_{*} = -0.186 \pm 0.021$, mass dependence parameter $\alpha$ is set free, and redshift dependence parameter is fixed to $\beta=0.666$ as in P16.
Intrinsic scatter for ACT scaling relation is also varied with a Gaussian prior on it as in {\it Planck} scaling relation, $\sigma_{\ln Y} = 0.173 \pm 0.023$. 
There is a possibility of covariance for common clusters, but we ignore it in this work since it is likely to be small for 11 clusters that overlap.
Following Section~\ref{ss:result_each}, we use two-dimensional likelihood functions for both P16 and Hi18 clusters.

\begin{figure*}
\centering
\includegraphics[width=\columnwidth]{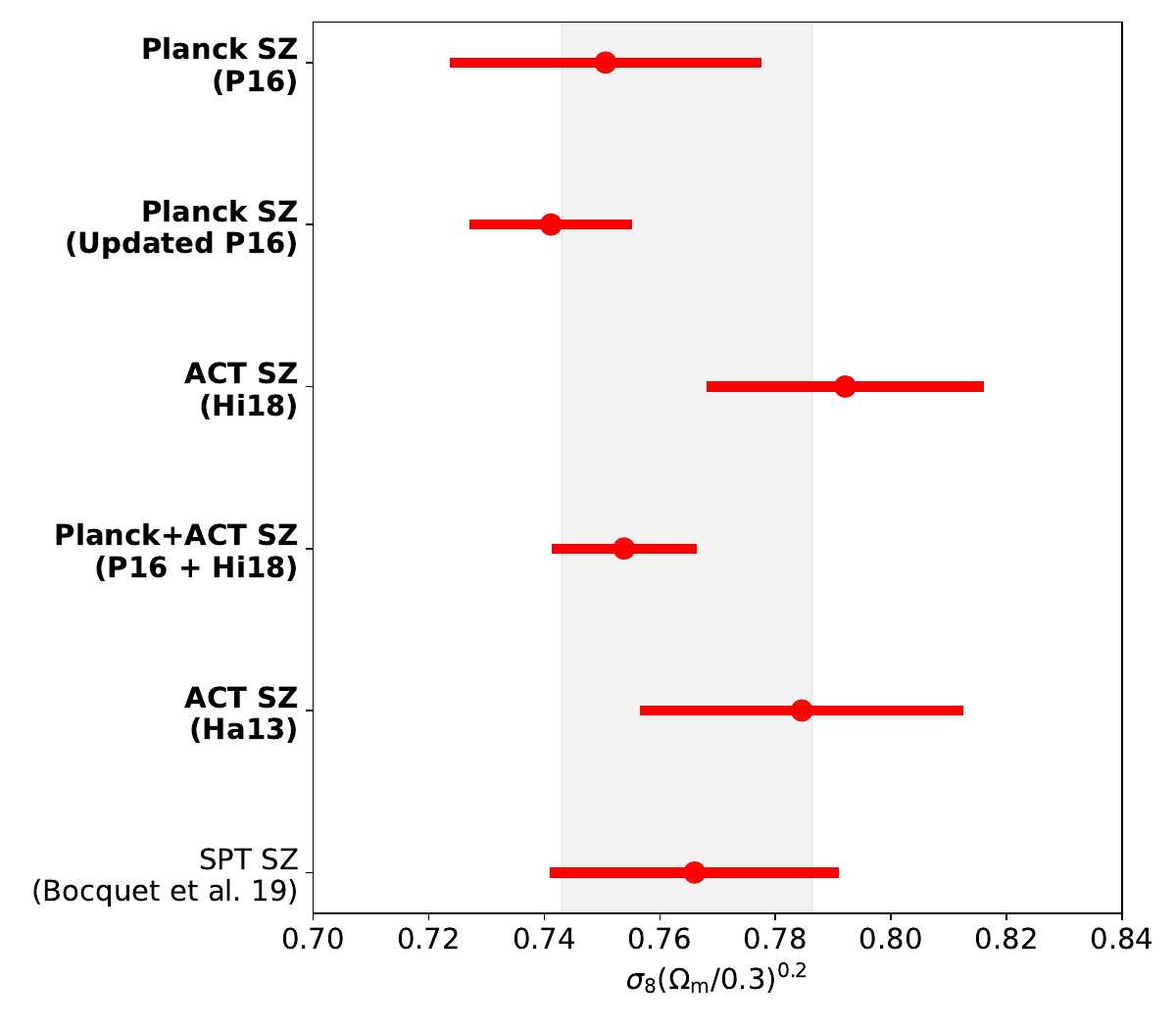}
\includegraphics[width=\columnwidth]{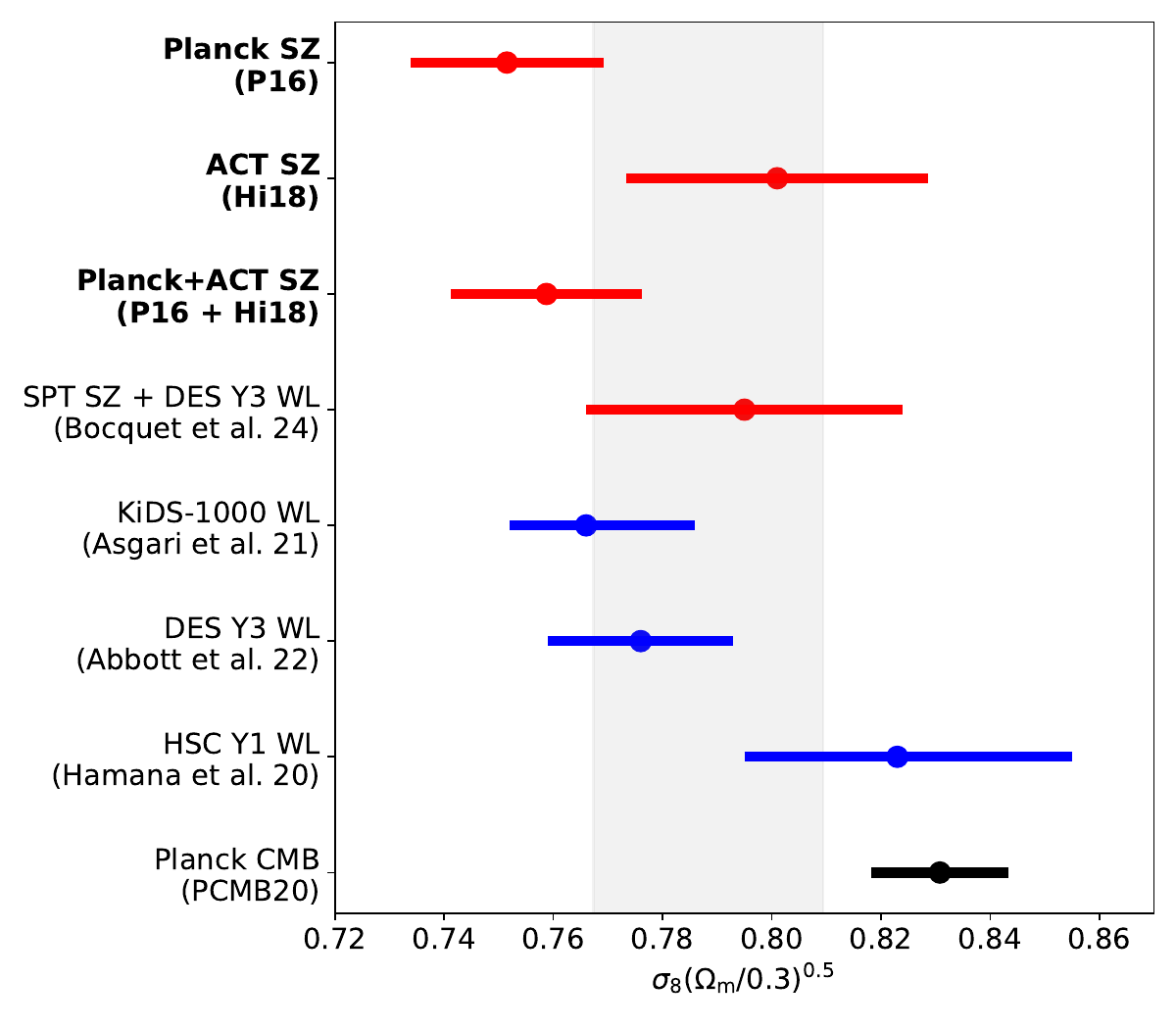}
\caption{Comparison of constraints in the $\Omega_{\rm m}-\sigma_8$ plane from different SZ datasets (left) and other cosmological probes (right). The measurements from this study are presented in bold. For SZ measurements we use the combination of $\sigma_8 \left( \Omega_{\rm m}/0.3 \right)^{0.2}$ which is standard in that context, while for the other cosmological probes, we compute  $\sigma_8 \left( \Omega_{\rm m}/0.3 \right)^{0.5}$. In the left panel, except for the SPT constraint which is directly from \protect\citealp{Bocquet2018}, all the other constraints are obtained from the binned likelihood computation presented in this paper using the {\it Planck} cluster model. SZ datasets are combined with the BAO and BBN data with a mass bias prior obtained from CCCP (\protect\citealp{Herbonnet2019cccp}) and shown in red colour. In the right panel, the constraints for other probes are taken from the literature (optical weak lensing measurements in blue and CMB measurements in black colour), while those from SZ datasets are from our binned likelihood and shown in red colour. The detail of each value is explained in subsection~\ref{sss:s8_constraints} and summarised in Table~\ref{tab:s8_comp}. The shaded area is given by the mean of the central value of the constraints and their uncertainties shown here.} 
\label{fig:s8_comp}
\end{figure*}
%

\subsection{Constraints in the $\Omega_{\rm m}-\sigma_8$ plane}\label{sss:s8_constraints}
Constraints on the $\sigma_8-\Omega_{\rm m}$ plane are the most sensitive ones that can be obtained from cluster measurements. 
Previously, the combination $\sigma_8 \left( \Omega_{\rm m}/0.27 \right)^{0.3}$ is used in {\it Planck} \citep{Planck2014xx}, ACT (Ha13), and SPT \citep{deHaan2016} and later the combination $\sigma_8 \left( \Omega_{\rm m}/0.31 \right)^{0.3}$ is used in P16. Here we present constraints on the combinations $\sigma_8 \left( \Omega_{\rm m}/0.3 \right)^{0.2}$ which is calculated in the recent SPT analysis \citep{Bocquet2018} for easier comparisons.

We have presented six different determinations of the combination $\sigma_8 \left(\Omega_{\rm m}/0.3 \right)^{0.2}$ in the left-hand panel of Figure~\ref{fig:s8_comp}. The measurements from this study are presented in bold in the Figure.
\begin{itemize}
    \item P16 using the priors detailed in \citealp{Planck2016xxiv}, in particular the BAO data, CCCP, and BBN priors which are different from those which we have detailed in Section~\ref{sec:external}. We find that $\sigma_8 \left(\Omega_{\rm m}/0.3 \right)^{0.2} = 0.751 \pm 0.027$. Note that \citealp{Planck2016xxiv} reports $\sigma_8 \left( \Omega_{\rm m}/0.31 \right)^{0.3} = 0.774 \pm 0.034$ and this is not directly comparable since different datasets for BAO and CCCP and a different definition of the parameter combination are used. 

   \item An updated result for P16 is $\sigma_8 \left(\Omega_{\rm m}/0.3 \right)^{0.2} = 0.741 \pm 0.014$. This uses the new information on the BAOs, the updated CCCP prior, and changes to the spectral index constraint included in the BBN prior. We note that the uncertainty is reduced by about a factor. 

    \item  Our new result is from using the Hi18 data with the Planck cluster model and priors from BAO, BBN, and CCCP. The value we find is $\sigma_8 \left( \Omega_{\rm m}/0.3 \right)^{0.2} = 0.792 \pm 0.024$. 
    
    \item A constraint can be obtained by combining P16 and Hi18 using the {\it Planck} cluster model yielding $\sigma_8\left(\Omega_{\rm m}/0.3 \right)^{0.2}=0.754 \pm 0.013$. The fact that they overlap but not completely overlap means that the combined uncertainty is reduced and the central value is shifted slightly upwards from the P16 value. We caution that this could be due to inadequacy in the cluster model as we will explore in the subsequent subsections.
    
   \item Our version of the constraint that would come from Ha13 presented in Figure~\ref{fig:act2013result} which is $\sigma_8 \left(\Omega_{\rm m}/0.3 \right)^{0.2} = 0.785 \pm 0.028$. This is recomputed from one-dimensional likelihood (only binned in redshift dimension) using BBN and $H_0$ priors taken from Ha13. 
    
    \item SPT SZ constraint taken directly from \citealp{Bocquet2018}, $\sigma_8 \left( \Omega_{\rm m}/0.3 \right)^{0.2} = 0.766 \pm 0.025$, which is obtained from combined cluster data and weak lensing measurement with X-ray information.
    
\end{itemize}
The shaded area is centered on the mean value of the constraint shown in the figure with the mean uncertainty of the given constraints. 
The basic picture that one might glean from this is that there is some evidence from a wide range of cluster-based probes that $\sigma_8 \left(\Omega_{\rm m}/0.3 \right)^{0.2}$ is less than 0.8, although we have already noted the CMBlens will lead to a lightly large value and significantly large uncertainty (see Figure~\ref{fig:sz_cccp_cmblensing}).

The parameter combination $\sigma_8 \left(\Omega_{\rm m}/0.3 \right)^{0.5}$ is commonly used for low-redshift cosmological probes other than clusters as it is a better-constrained parameter by weak lensing analyses. 
In the right panel of Figure~\ref{fig:s8_comp}, we compare a determination of $\sigma_8 \left(\Omega_{\rm m}/0.3 \right)^{0.5}$ from different cosmological probes.

\begin{itemize}
\item First, we computed this combination for the updated P16, the Hi18, and combined P16/Hi18 analyses. 
We find that P16 gives $\sigma_8 \left(\Omega_{\rm m}/0.3 \right)^{0.5} = 0.751 \pm 0.018$, Hi18 gives $\sigma_8 \left(\Omega_{\rm m}/0.3 \right)^{0.5} = 0.801 \pm 0.028$, and the combined P16/Hi18 gives $\sigma_8 \left( \Omega_{\rm m}/0.3 \right)^{0.5} = 0.759 \pm 0.018$. 
These are presented in red in the right-hand panel as they are in the left-hand panel. 
We note that the uncertainties on this quantity and slightly larger than those for $\sigma
\left(\Omega_{\rm m}/0.3 \right)^{0.2}$ indicating that the latter is a more optimal combination for clusters.
Also, we have included the very recent constraint from \citealp{bocquet2024arXiv240102075B}: $\sigma_8 \left(\Omega_{\rm m}/0.3 \right)^{0.5} = 0.795 \pm 0.029$ which is obtained from SPT clusters using DES weak lensing mass calibration.
\item We have compared these SZ determinations with other low redshift probes including the measurements from cosmic shear and galaxy clustering which are shown in blue. 
KiDS-1000 constraint from \citealp{KiDS2021}: $\sigma_8\left(\Omega_{\rm m}/0.3 \right)^{0.5}=0.766_{-0.020}^{+0.014}$ from cosmic shear measurements by KiDS and the spectroscopic galaxy clustering from BOSS.
DES Y3 constraint from \citealp{DES2022}: $\sigma_8 \left(\Omega_{\rm m}/0.3 \right)^{0.5}=0.776 \pm 0.017$, obtained by combining three two-point correlation functions (cosmic shear, galaxy clustering and the cross-correlation between them).
Subaru HSC Y3 constraint from \citealp{HSCY3_dalal2023}: $\sigma_8 \left(\Omega_{\rm m}/0.3 \right)^{0.5} = 0.776_{-0.033}^{+0.032}$ by using cosmic shear power spectra.

\item In addition we have included the constraint in black from PCMB20, $\sigma_8 \left(\Omega_{\rm m}/0.3 \right)^{0.5} = 0.831 \pm 0.013$.
Although the picture is not entirely clear there does appear to be a discrepancy between the low redshift probes and the CMB \citep[see, e.g.,][]{Battye:2014qga}.

\end{itemize}


%
\begin{table}
    \centering
    \begin{tabular}{l l}
         \hline \hline
         SZ datasets & Constraints \\ \hline
         P16 + CCCP (\cite{Hoekstra2015}) & 
         $\sigma_8 \left( \Omega_{\rm m}/0.3 \right)^{0.2}=0.751 \pm 0.027$ \\
         P16 + CCCP & 
         $\sigma_8 \left( \Omega_{\rm m}/0.3 \right)^{0.2}=0.741 \pm 0.014$ \\
         Ha13 + CCCP  & 
         $\sigma_8 \left( \Omega_{\rm m}/0.3 \right)^{0.2}=0.785 \pm 0.028$ \\          
         Hi18 + CCCP & 
         $\sigma_8 \left( \Omega_{\rm m}/0.3 \right)^{0.2}=0.792 \pm 0.024$ \\ 
         P16 + Hi18 + CCCP & 
         $\sigma_8 \left( \Omega_{\rm m}/0.3 \right)^{0.2}=0.754 \pm 0.013$ \\  \hline 
         P16 + CCCP  & 
         $\sigma_8 \left( \Omega_{\rm m}/0.3 \right)^{0.5}=0.751 \pm 0.018$ \\   
         Hi18 + CCCP & 
         $\sigma_8 \left( \Omega_{\rm m}/0.3 \right)^{0.5}=0.801 \pm 0.028$ \\   
         P16 + Hi18 + CCCP & 
         $\sigma_8 \left( \Omega_{\rm m}/0.3 \right)^{0.5}=0.759 \pm 0.018$ \\          
         \hline
    \end{tabular}
    \caption{Summary of constraints obtained in this work in $\Omega_{\rm m}-\sigma_8$ plane from different SZ datasets in Figure~\ref{fig:s8_comp}. All the SZ datasets include the external data of BAO and BBN. The latest measurement of mass bias from CCCP (\protect\cite{Herbonnet2019cccp}) is used, unless stated otherwise.}
    \label{tab:s8_comp}
\end{table}

\subsection{Constraining cluster properties: $\alpha$ and $\beta$}\label{ss:beta_varied}
%
%
\begin{figure*}
    \centering
    \includegraphics[width=0.90\columnwidth]{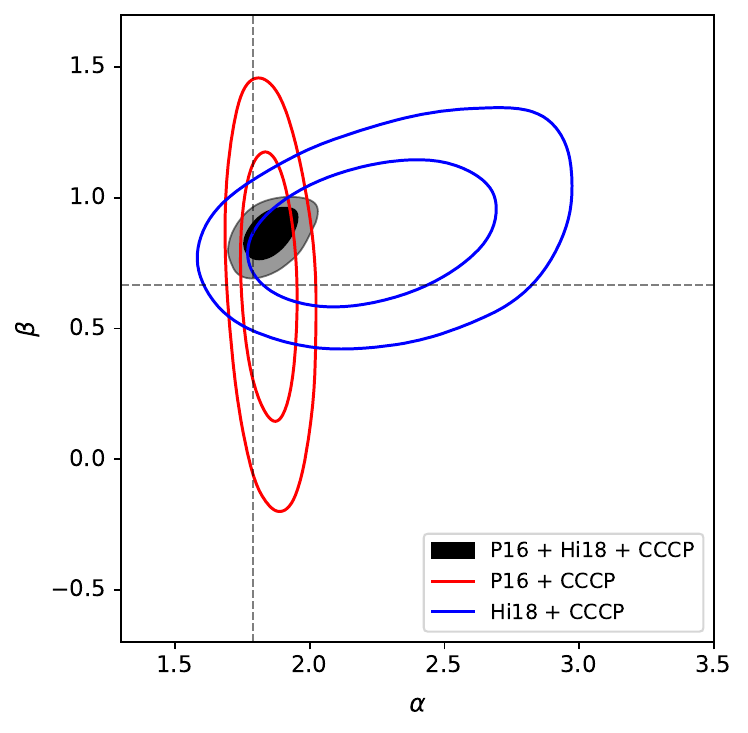}
    \includegraphics[width=0.89\columnwidth]{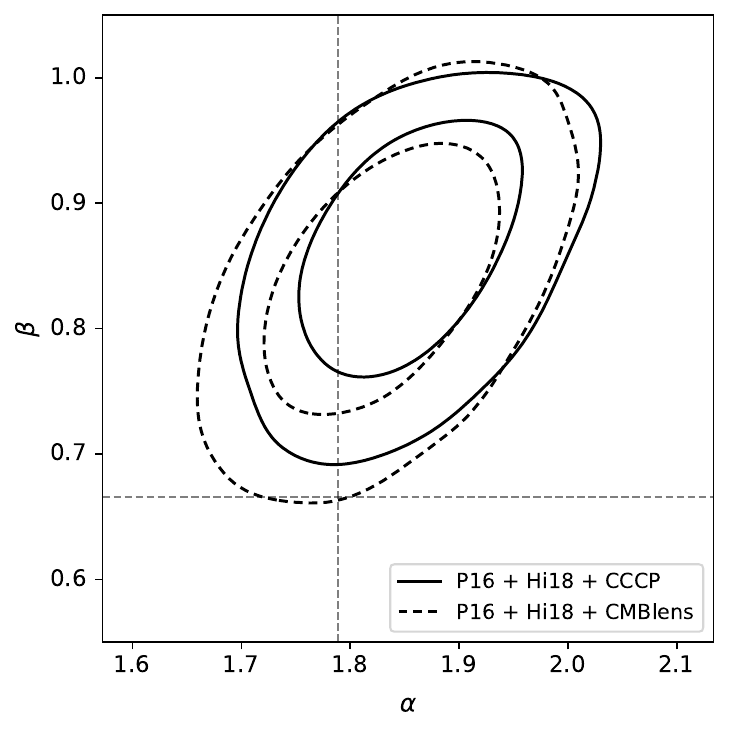}
    \caption{Comparison of constraints on mass slope $\alpha$ and redshift slope $\beta$. Our baseline scenario has been to allow $\alpha$ to vary, while $\beta$ has been fixed to the standard value $2/3$ as advocated in P16. In the left panel, the black filled contour is obtained using the P16/Hi18 combined dataset, the red contour is from using P16 only, and the blue contour is from using Hi18 only. We have imposed priors for BAO data, BBN, and CCCP prior. P16 alone does not constrain $\beta$ since the clusters have a narrow range of low redshifts whereas Hi18 can constrain $\beta$ relatively well due to the wider redshift range probed, even though the sample size is significantly lower. Conversely, P16 constrains $\alpha$ much since the sample is larger and more finely binned in the SNR dimension. The right panel shows the comparison of the results from different mass bias priors imposed. The black solid line is using the CCCP prior and the black dashed line is using the CMBlens prior. Vertical and horizontal dashed lines in both panels show the central value of Gaussian prior used for $\alpha$ and a standard value used for $\beta$ in P16.}
    \label{fig:beta_varied}
\end{figure*}
%
%
The assumption we have made in order to combine the P16 and Hi18 data is that the two sets of clusters have the same scaling relation, but we have also noted that for example using the PCMB20 data to fix the cosmological parameters that the Hi18 clusters prefer a slightly larger value of $1-b$. 
This could be due to the redshift evolution of the scaling relation and the associated parameter is the power low index $\beta$, which is typically assumed to be $2/3$ in the cluster model.
We test this hypothesis by allowing $\beta$ to vary in the scaling relation while fixing the bias using the CCCP and CMBlens priors. 
Since there is a degeneracy between the normalization of the scaling relation and the mass bias this can be equally thought of as the time evolution of the scaling relation, or evolution in the mass bias.

Before combining the P16 and Hi18 data, we investigate the impact of $\beta$ variation on parameter combination of $\sigma_8 \left(\Omega_{\rm m}/0.3 \right)^{0.2}$ for each dataset.
We find $\sigma_8 \left(\Omega_{\rm m}/0.3 \right)^{0.2} = 0.741 \pm 0.019$ for P16 and $\sigma_8 \left(\Omega_{\rm m}/0.3 \right)^{0.2} = 0.766 \pm 0.034$ for Hi18.
P16 finds the same central value but with a larger uncertainty whereas the central value for Hi18 shifts apparently on the lower side. 
It also shows that allowing $\beta$ to vary reduces $\sim3\sigma$ difference of constraints on $\sigma_8 \left(\Omega_{\rm m}/0.3 \right)^{0.2}$ between P16 and Hi18 to $1\sigma$ level. 
When combining two datasets with a varied $\beta$ parameter, we find a reduced $\chi^2$ value from 230 to 224, which confirms that this gives a better fit.

In Figure~\ref{fig:beta_varied} we present the posterior distributions of scaling parameters $\alpha$ and $\beta$ from the combined cluster data sets of {\it Planck} and ACT with BAO and BBN priors. 
In the left-hand panel we use the CCCP prior and how results for using P16, Hi18, and the combination of P16 and Hi18 combined. 
It is clear from this figure that P16 constrains $\alpha$ rather well compared to Hi18, while the converse is true for the parameter $\beta$ with Hi18 constraining it well and there is a wide range of values of $\beta$ which are compatible with the P16 clusters.
When they are combined one finds accurate determinations of both $\alpha = 1.855 \pm 0.068$ and $\beta = 0.861 \pm 0.067$. 
This can be easily explained by realising that the P16 data has a number of SNR bins - with SNR being a proxy for mass - constraining $\alpha$ and the Hi18 covers a much wider range of redshifts allowing a constraint on $\beta$. 
We note that both these values are not compatible with the standard values $\alpha=5/3$ and $\beta=2/3$ predicted, for example, by simple isothermal self-similar cluster models. 
In the right-hand panel of figure~\ref{fig:beta_varied} we demonstrate that this constraint is relatively insensitive to the choice of the mass bias prior but comparing the contours for the constraints in the $\alpha-\beta$ plane for P16 and Hi18.
The marginalised constraints on $\alpha$ and $\beta$ are presented in Table~\ref{tab:beta_varied} for the two mass bias priors.


%
\begin{table}
    \centering
    \begin{tabular}{c c c}
         \hline \hline
         Parameter & P16 + Hi18 + CCCP & P16 + Hi18 + CMBlens  \\ \hline
         $\alpha$ & $1.855 \pm 0.068$ & $1.830 \pm 0.071$ \\
         $\beta$ & $0.861 \pm 0.067$ & $0.839 \pm 0.072$ \\
         \hline
    \end{tabular}
    \caption{Constraints on mass dependence $\alpha$ and redshift dependence $\beta$ in scaling relation when cluster data are combined. Corresponding contours are in the right panel of Figure~\ref{fig:beta_varied}.}
    \label{tab:beta_varied}
\end{table}
%
%
%

\subsection{Constraining equation of state of dark energy $w$}
\label{ss:wwmeasure}

\begin{figure}
    \centering
    \includegraphics[width=0.9\columnwidth]{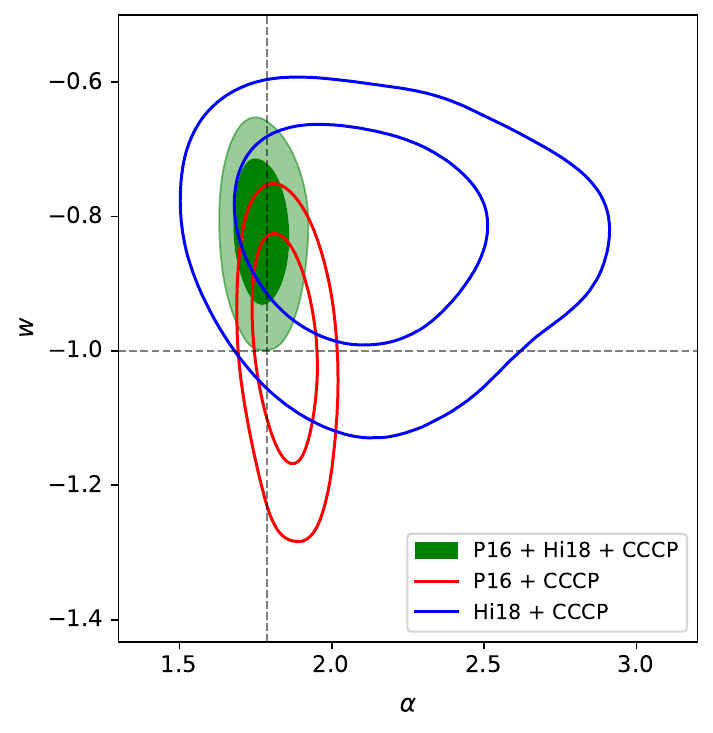}
    \caption{Comparison of constraints on the equation of state of dark energy $w$ and the mass dependence scaling parameter $\alpha$ when $w$ is varied free and the redshift dependence parameter $\beta$ in scaling relation is fixed. Green-filled contour shows the constraint from using the P16/Hi18 combined dataset. The red contour is from using P16 alone and the blue is from using Hi18 alone. Gray dashed lines show the standard value for $w$ and $\alpha$.}
    \label{fig:w_varied}
\end{figure}

In the previous section, we showed that the inclusion of redshift dependence in the cluster scaling relations can bring the P16 and Hi18 analyses into alignment. 
One might wonder whether this effect could also be due to a modification in the evolution of the Universe. 
In order to probe this possibility we allow the dark energy equation of state parameter $w$ to vary. 
Changing the history of the expansion of the Universe modifies both the growth rate of structures and the distances to objects which leads to changes in the cluster abundance \citep[see, e.g.,][]{Battye:2003bm}. 
In the preceding analyses we have fixed $w=-1$, but now we let $w$ vary in a range of $[-4, 0]$ and fix $\beta$ to $2/3$, and the constraints are presented in Figure~\ref{fig:w_varied} using the BAO, BBN, and CCCP priors. 
The results are shown on a plane of the dark energy parameter $w$ and the mass dependence parameter $\alpha$ in scaling relation in order to compare to the behaviour in the $\alpha-\beta$ plane in Figure~\ref{fig:beta_varied}. 
The green filled contour is from P16 and Hi18 combined dataset, the red contour is for P16 alone, and the blue is for Hi18 alone. 
We do see a similar, but possibly not quite so striking effect of including $w$ instead of $\beta$. 
When using the combined data set, we deduce tighter constraints on both $w$ and $\alpha$, which are $w=-0.823 \pm 0.072$ and $\alpha=1.770 \pm 0.061$.
Similarly, as in Section~\ref{ss:beta_varied}, by letting $w$ free, we find a reduced difference (at $1\sigma$ level) in constraint on $\sigma_8 \left(\Omega_{\rm m}/0.3 \right)^{0.2}$ for individual dataset.
When both datasets are combined, the value of $\chi^2$ reduces from 230 to 225 as well.

%

\begin{figure*}
    \centering
    \includegraphics[width=0.7\textwidth]{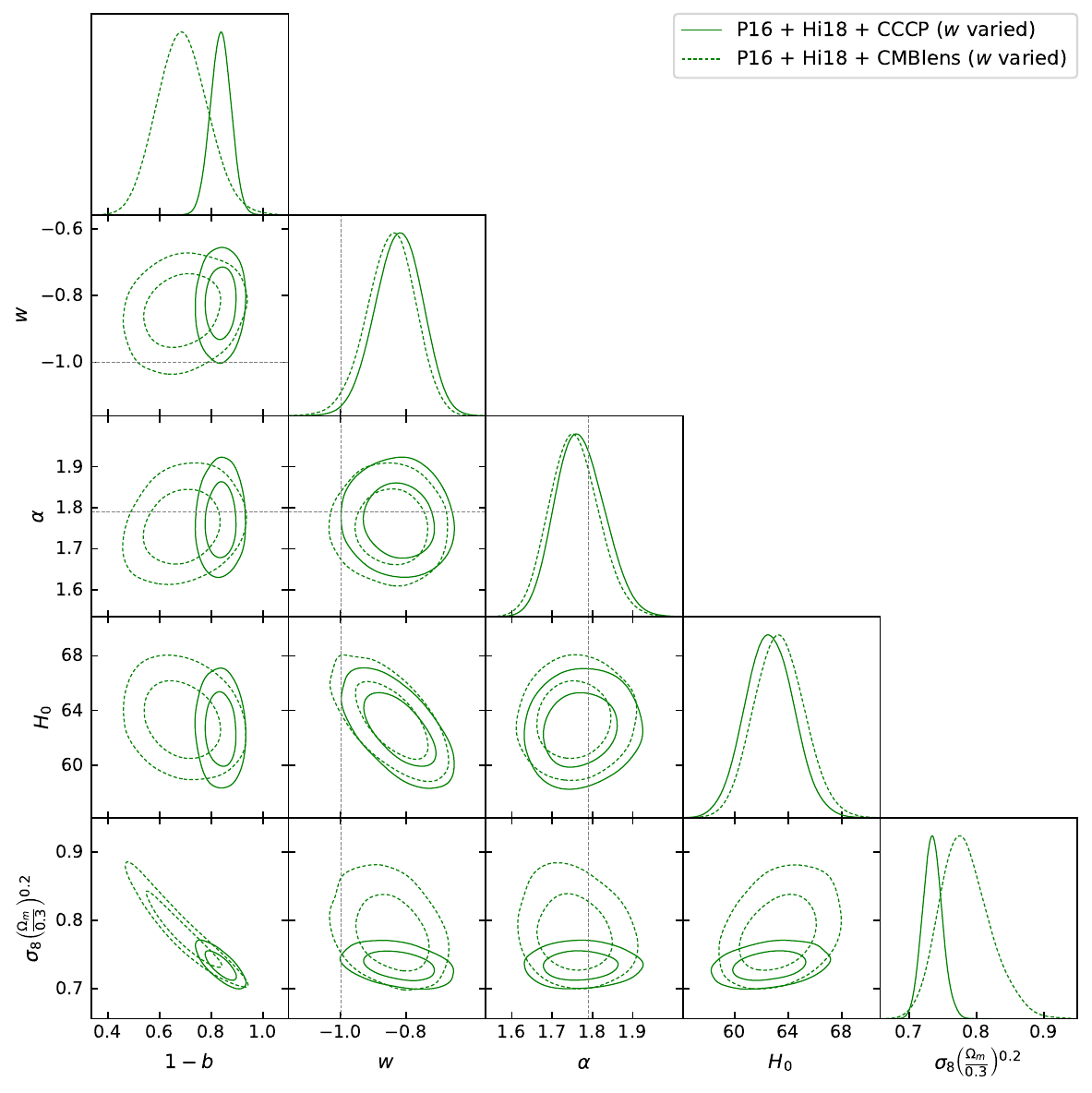}
    \caption{Comparison of constraints on cosmological and cluster scaling parameters when the equation of state of dark energy $w$ is varied free and the redshift dependence parameter $\beta$ in scaling relation is fixed. Green solid line contours represent the results from imposing the CCCP prior and green dashed line contours represent the results from the CMBlens prior. Note that the constraints on $w$ and $H_0$ appear to be relatively independent of the mass bias prior, whereas that on $\sigma_8\left({\Omega_{\rm m}/0.3}\right)^{0.2}$ significant affected by the prior.}
    \label{fig:w_varied_together}
\end{figure*}

Figure~\ref{fig:w_varied_together} illustrates how the cosmological constraints change depending on the choice of a prior on the mass bias using the P16/Hi18 combination with the BAO and BBN priors.
We see that the different choice of priors, CCCP and CMBlens, does not have a significant impact on the equation of state of dark energy equation of state parameter, $w$, and the Hubble constant, $H_0$, even though the two priors are centred at differing values and have different levels of uncertainty. 
However, we see that the constraints on combination $\sigma_8 \left(\Omega_{\rm m}/0.3 \right)^{0.2}$ are sensitive to the central value and the uncertainty of the mass bias; this is a further explicit illustration of the effect already in previous sections. 
For this particular, data/model combination we find that  $\sigma_8 \left(\Omega_{\rm m}/0.3 \right)^{0.2} = 0.734 \pm 0.014$ for the CCCP prior and  $\sigma_8 \left(\Omega_{\rm m}/0.3 \right)^{0.2} = 0.783 \pm 0.038$ for the CMBlens prior.
%
%

\subsection{Constraining both $\beta$ and $w$ from cluster data}
\begin{figure}
    \centering
    \includegraphics[width=0.9\columnwidth]{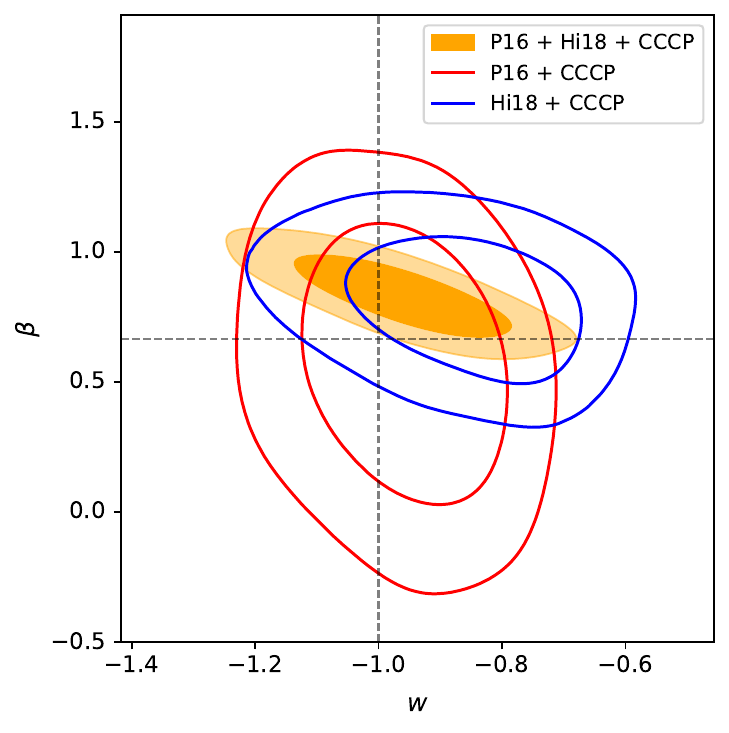}
    \caption{Comparison of constraints when both equations of state of dark energy $w$ and $\beta$ are allowed to be varied. Orange-filled contour is when P16 and Hi18 datasets are combined, the red unfilled contour is for the P16 data alone, and the blue empty contour is for Hi18 alone. All three cases include BAO, BBN data, and CCCP prior to mass bias.} 
    \label{fig:beta_and_w_varied}
\end{figure}

In the two previous sections, we have seen that $\beta$ and $w$ can alleviate the relatively minor, yet apparent discrepancy between P16 and Hi18 cluster samples using the same {\it Planck} cluster model and we have argued that they do this in a similar way by introducing some redshift dependence into the model, be it in the scaling relation for the case of $\beta$ and in the cosmological evolution for $w$. 
This suggests that the two should be correlated and indeed Figure~\ref{fig:beta_and_w_varied} illustrates this for the case of the combined P16/Hi18 cluster sample with BAO, BBN, and CCCP priors. 
We have also included the individual cases P16 and Hi18 to show the impact combining them has on the correlation. 
For the combined dataset, we find $\beta = 0.833 \pm 0.104$ and $w = -0.963 \pm 0.114$. 
As both these parameters are connected to redshift evolution, the constraint from Hi18 is stronger even though the sample size is smaller.
%
\section{Conclusions}
\label{ss:conc}
In this work, we have updated the P16 cluster count analysis for the cosmological parameter from \citealp{Planck2016xxiv} and extended it to include the Hi18 cluster counts based on \citealp{Hilton2017} including the latest weak lensing mass measurements. 
We take a binned likelihood approach and we are able to combine the two. 
The original MCMC sampler used for computation of likelihood has been switched from {\tt CosmoMC} \citep{Lewis2002} written in {\tt Fortran90} to {\tt COBAYA} \citep{cobaya2021} written in {\tt Python}. 

First, we were able to compare the cosmological constraints obtained using a binned likelihood method with the cosmological analysis from Ha13 and we found that they are in good agreement. 
This gave us confidence that our likelihood approach was sound.

We showed the observable-mass scaling relations used by P16 and Hi18 can be related, essentially by connecting the peak Compton-y parameter $y_0$ and integrated Compton-y parameter $Y_{500}$ using the UPP from A10. 
The specific scaling parameters used by P16 are calibrated from X-ray observations of Planck clusters, whereas those used in Ha13 and Hi18 are analytically derived from the UPP model.
Ultimately, they are not in exact agreement with each other. 
We took the {\it Planck} cluster model as the baseline for our combined analysis making the assumption that the clusters are from the same universal population.

Using two updated mass bias determinations from weak lensing, we report the constraints $\sigma_8 \left(\Omega_{\rm m}/0.3 \right)^{0.2}$ from P16, Hi18, and the P16/Hi18 combined dataset. 
All of these are typically smaller than the same quantity derived from PCMB20 observations within the framework of a flat $\Lambda$CDM model.

We find that there is a slight mismatch between the bias values derived assuming the same cosmological parameter defined by the PCMB20 analysis possibly indicating some evolution either in the cluster scaling relation quantified by the parameter $\beta$ - which is probably not unexpected and also most likely - or the expansion history of the Universe. The constraints on nuisance parameters, $\alpha$ and $\beta$, do not seem to be strongly dependent on either the central values or the size of uncertainties of the mass bias prior we impose.


Recent work from \citealp{Salvati:2021gkt} shows the results by combining {\it Planck} and SPT cluster samples.
The angular resolution of SPT is similar to that of ACT (Ha13 and Hi18), but SPT covers a larger survey area (2500 deg$^2$), and SPT clusters are extracted from the multi-frequency matched filter at frequencies of 95 and 150 GHz. 
The SPT cluster likelihood function is based on SZ cluster data together with X-ray and weak lensing measurements without using priors from external data. 
Individual likelihoods for {\it Planck} and SPT are computed for each cluster model and then combined with a modification on the {\it Planck} likelihood to take into account the overlapping area of the observed sky of {\it Planck} and SPT. 
In the $\nu\Lambda$CDM scenario, they find the mass slope of $\alpha = 1.49_{-0.10}^{+0.07}$ and the mass bias of $1-b = 0.69_{-0.14}^{+0.07}$ in {\it Planck} scaling relation, respectively.
They also explore the possibility of modeling the mass bias as a function of mass, redshift, and detection noise and provide a new catalogue of {\it Planck} cluster masses.
They find the mass bias to have an increasing trend with redshift. 
Also, more recently, \citealp{bocquet2024arXiv240102075B} shows the cosmological constraints from combined SPT sample (SPT-SZ, SPTpol ECS, and SPTpol 500d) using weak lensing data from the DES Y3 and the Hubble Space Telescope. They find $\sigma_8 \left(\Omega_{\rm m}/0.3 \right)^{0.25} = 0.805 \pm 0.016$ for a flat $\Lambda$CDM cosmology, which is consistent with, but tighter than, their previous results \citep{Bocquet2018}. They also find a good agreement with the {\it Planck} CMB results \citep{Planck2018lensing}.
Recently, \citealp{qu2023atacama} presents the cosmological constraints derived from ACT CMB lensing power spectrum. They find $\sigma_8 \left(\Omega_{\rm m}/0.3 \right)^{0.25} = 0.818 \pm 0.022$ from ACT DR6 lensing power spectrum alone, which is consistent with both the previous ACT \citep{act2020aiola} and the {\it Planck} CMB results \citep{Planck2018lensing}.

Although the size of the ACT cosmological sample used in this analysis is only 59, this work lays the groundwork for using binned likelihood with the ACT data. It would be interesting to see how much we can benefit from a much larger sample from the latest ACT SZ cluster catalogue (\citealp{Hilton2021}, ACT DR5), which has $\sim$4000 cluster candidates (SNR > 4).
Once ACT DR6 SZ cluster dataset becomes available it will be interesting to revisit the joint analysis of {\it Planck} and ACT (and potentially SPT) using cluster by cluster likelihood such as the one implemented in \texttt{cosmocnc}\footnote{\href{https://github.com/inigozubeldia/cosmocnc}{https://github.com/inigozubeldia/cosmocnc}} (\citealp{Zubeldia:2024lke}).



\section*{Acknowledgements}

We would like to thank Nicholas Battaglia, Jens Chluba, Matt Hilton, Mat Madhavacheril, Jean-Baptiste Melin, Laura Salvati, and Inigo Zubeldia for discussions and/or comments on this work.
\\
EL acknowledges support from UK Science and Technology Facilities Council (STFC) under grant ST/P006795/1.
BB acknowledges support from the European Research Council (ERC) under the European Union’s Horizon 2020 research and innovation programme (Grant agreement No. 851274). \\
Some of the results in this paper have been derived using {\tt Matplotlib} \citep{Hunter:2007} library and {\tt GetDist} \citep{Lewis2019} software package. 





\bibliographystyle{mnras}
\bibliography{main.bib}








\bsp	
\label{lastpage}
\end{document}